\documentclass{mn2e}

\usepackage{epsfig}

\def\hvect#1{{\hat{\mathbfit{#1}}}}

\title{All-sky component separation for the Planck mission}

\author[]{V.~Stolyarov$^{1,2}$, M.P.~Hobson$^2$, M.A.J.~Ashdown$^2$, 
A.N.~Lasenby$^2$\\
$^1$ Institute of Astronomy, Madingley Road, Cambridge CB3 0HA, UK\\
$^2$ Astrophysics Group, Cavendish Laboratory, Madingley Road, 
Cambridge CB3 0HE, UK}

\date{Accepted ???. Received ???; in original form \today}

\pagerange{\pageref{firstpage}--\pageref{lastpage}}
\pubyear{2001}

\begin{document}

\maketitle

\label{firstpage}

\begin{abstract}
An harmonic-space maximum-entropy method (MEM) is presented for separating the
emission from different physical components in all-sky 
observations by the forthcoming Planck satellite.
The analysis is performed at full Planck resolution, with a pixel size 
of 1.7 arcmin, which corresponds to $\ell_{\rm max} \approx 6000$.
The simulated Planck data include
emission from the CMB, the kinetic and thermal Sunyaev-Zel'dovich (SZ) effects
from galaxy clusters, as well as Galactic dust, free-free and
synchrotron emission. 
Our simulations also assume homogeneous, uncorrelated
pixel noise, although this is not a requirement of the method.
We find that the MEM technique produces faithful 
reconstructions of the main input components over the whole sky, without 
the need to perform a Galactic cut. The CMB power spectrum
is accurately recovered up to $\ell \approx 2000$.
The algorithm is parallelised so that
the entire reconstruction can be performed in $\sim$ 6 hr using 30 R10000
processors on an SGI Origin 2000 supercomputer and 
requires 14 Gb of RAM.
\end{abstract}

\begin{keywords}
methods -- data analysis -- techniques: image processing -- cosmic
microwave background.
\end{keywords}

\section{Introduction}
\label{intro}

Two new cosmic microwave background (CMB) satellite
missions are currently in preparation.  The NASA MAP satellite is
expected to be launched in late 2001, followed by the ESA Planck
Surveyor in 2007.  Both missions will provide
detailed all-sky maps at a number of observing frequencies.  The main
aim of these new satellite projects is to obtain an accurate map of
CMB anisotropies over the whole sky and produce a definitive
measurement of the CMB power spectrum.  This should allow tight
constraints to be placed on fundamental cosmological parameters and
distinguish between competing theories of structure formation.

The maps produced by these satellites will, however, contain
contributions from various foreground components, most notably
Galactic dust, free-free and synchrotron emission as well as the
kinetic and thermal Sunyaev-Zel'dovich (SZ) effects from galaxy clusters. In addition,
significant contamination from extragalactic point sources is also
expected. It is therefore clear that in order to obtain maps of the
CMB anisotropies alone, it is necessary to separate the emission due
to these various components. Traditional methods for performing the
separation include singular-valued decompostion and Wiener filtering
(Bouchet, Gispert \& Puget 1996; Tegmark \& Efstathiou 1996), although a recent
preliminary application of neural-networks to this problem appears
promising (Baccigalupi et al. 2000). A review of traditional foreground
separation techniques is given by Bouchet \& Gispert (1999).

In a previous paper (Hobson et al. 1998); hereafter Paper I), a
separation was performed on simulated Planck Surveyor observations of a $10\times
10$ deg$^2$ field (see Bouchet et al. 1997; Gispert \& Bouchet 1997),
using a non-linear Fourier-space maximum-entropy method (MEM) that
reduces to traditional Wiener filtering in the absence of non-Gaussian
signals.  It was found that faithful reconstructions may be produced
not only of the CMB anisotropies but also of the Galactic components
and the thermal SZ effect from massive clusters. It was also shown
that the MEM technique outperformed standard Wiener filtering,
particularly in reconstructing highly non-Gaussian components such as
the SZ effects.  An application of the MEM component separation
technique to simulated MAP data of a $10\times
10$ deg$^2$ field was presented by Jones, Hobson \& Lasenby (1999)

In Hobson et al. (1999), the basic MEM algorithm was
extended to identify and remove contamination from extragalactic point
sources, and has been further refined by Vielva et al. (2001) by
combining it with a mexican hat wavelet filtering technique.

A great advantage of forthcoming satellite missions, however, is the
prospect of obtaining all-sky maps at each observing frequency, from
which one would hope to reconstruct maps of the emission from each
physical component over the whole sky simultaneously, rather than just
in small patches. Indeed, Prunet et al. (2001) recently presented the
first application of the Wiener filter component separation algorithm
to simulated all-sky data from the MAP experiment. Their analysis was
performed up to $\ell_{\rm max} = 512$ and included contributions from
the three dominant astrophysical components expected in the five MAP
frequency channels, namely CMB and Galactic dust and synchrotron
emission.

It is, however, the Planck mission that provides the greatest
challenge for performing a component separation.  Planck will map the
whole sky in 10 frequency bands from 30 to 857 GHz and at angular
resolutions ranging from 33 to 4.5 arcmin.  To ensure Nyquist sampling
of the sky emission, the basic pixel size for Planck reconstructions
is $\sim 2$ arcmin, which corresponds to $\sim 50$ million pixels in
each all-sky map and maximum multipole $\ell_{\rm max} \approx
6000$. However, the problem is not simply one of computational
complexity. Owing to its greater frequency coverage and higher angular
resolution, Planck is sensitive to the full range of astrophysical
foreground outlined above, and thus provides an opportunity to map
each of these components individually. In particular the highly
non-Gaussian SZ effects from galaxy clusters are of considerable
cosmological interest.

In this paper, we therefore extend the Fourier-space MEM algorithm to
perform reconstructions on the whole celestial sphere.  Since the
algorithm is inherently computationally fast and efficient, and,
moreover, can be straightforwardly parallelised to take advantage of
existing supercomputing facilities, we are able to perform
reconstructions to full Planck resolution.  As well as the vastly
increased computational burden over the analysis presented in Paper I,
when performing all-sky reconstructions we also encounter the problem
of highly inhomogeneous emission due to Galaxy. The latter leads to
foreground maps with a very large dynamic range. Nevertheless, we show
that presence of bright emission from the Galactic plane does not
severely affect the accuracy of the reconstructions obtained, and that
it is {\em not} necessary to impose a Galactic cut on the data prior
to the component separation analysis. We note, however, that a Galactic
cut can still be performed in order to prevent contamination from poorly
modelled strong Galactic emission regions.

\section{Model of the microwave sky}
\label{model}

To create simulated Planck observations, we must first build a
plausible model of the emission over the whole sky at each Planck
frequency.  We assume the main contributions to this emission are from
the primordial CMB, the kinetic and thermal SZ effects from galaxy
clusters, and the Galactic dust, synchrotron and free-free components.
In this paper, we will assume that extragalactic point sources may be
removed earlier in the analysis using the satellite observations
themselves together with existing surveys, or by applying the joint
MEM and mexican hat wavelet technique discussed in Vielva et
al. (2001) to the data map at each frequency. The generalisation of
this joint scheme to all-sky observations and using all
the frequencies simultaneously will be presented in a
forthcoming paper.

Following the notation of Paper I, we choose to work in units of
equivalent CMB thermodynamic temperature. 
If $\Delta I(\bmath{x}) $ is the fluctuation in the
specific intensity at some frequency $\nu$ in the direction
$\bmath{x}$, then the corresponding fluctuation in the equivalent 
CMB thermodynamic temperature is given by
\[
\Delta T(\bmath{x},\nu) 
= \frac{\Delta I(\bmath{x},\nu)}
{\left.\partial B(\nu,T)/\partial T\right|_{T=T_0}},
\]
where $B(\nu,T)$ is the Planck function and $T_0 = 2.726$ K is the
temperature of the CMB (Mather et al. 1994).
As in paper I, we also assume that the contribution
from the $p$th physical component can be factorised into a spatial
template $s_p(\hvect{x})$ at reference frequency $\nu_0$ and a
frequency dependence $f_p(\nu)$, so that
\begin{equation}
\Delta T(\hvect{x},\nu) = \sum_{p=1}^{n_c} \Delta T_p(\hvect{x},\nu) =
\sum_{p=1}^{n_c} f_p(\nu)s_p(\hvect{x}).
\label{sumcomps}
\end{equation}
We take as our reference frequency $\nu_0$ = 300 GHz, and normalise
the frequency behaviour so that $f_p(\nu_0)$ = 1 for all the physical
components.

It is clear that (\ref{sumcomps}) represents the somewhat idealised
case in which the spectral index of each physical component does not
vary with direction on the sky.  While this assumption is valid for
the CMB and the two SZ effects, it is unlikely to hold for the
Galactic components.  This was not a severe concern in Paper I, since
the analysis there was restricted to a 10 $\times$ 10 deg$^2$ patch of
sky over which the spectral indices of the Galactic components might
reasonably be expected not to vary significantly (particularly in
regions of high Galactic latitude). For an all-sky reconstruction,
however, the assumption almost certainly does not hold, even
approximately. Nevertheless, as mentioned in Paper I and investigated
further in Jones et al. (1999), it is possible for the separation
algorithm to accommodate varying spectral indices by including extra
components in the reconstruction. For example, if we assume that the
frequency dependence of the synchrotron emission is of the form $I
\propto \nu^{-\beta}$, with $\beta=-0.7\pm 0.2$, we simply include two
synchrotron channels, one with $\beta=-0.5$ and one with $\beta=-0.9$,
or even with intermediate values, and afterwards sum over these
channels to obtain the reconstructed synchrotron map. Alternatively,
one can consider deviations from the mean spectrum as just another
template to be recovered with a modified spectral behaviour as
obtained by linearising the frequency dependence of the intensity with
these deviations (Bouchet et al. 1996). We will pursue these ideas in
a forthcoming paper, but for now we consider only the idealised
situation represented by (\ref{sumcomps}).

\begin{figure*}
\begin{center}
\centerline{
\epsfig{file=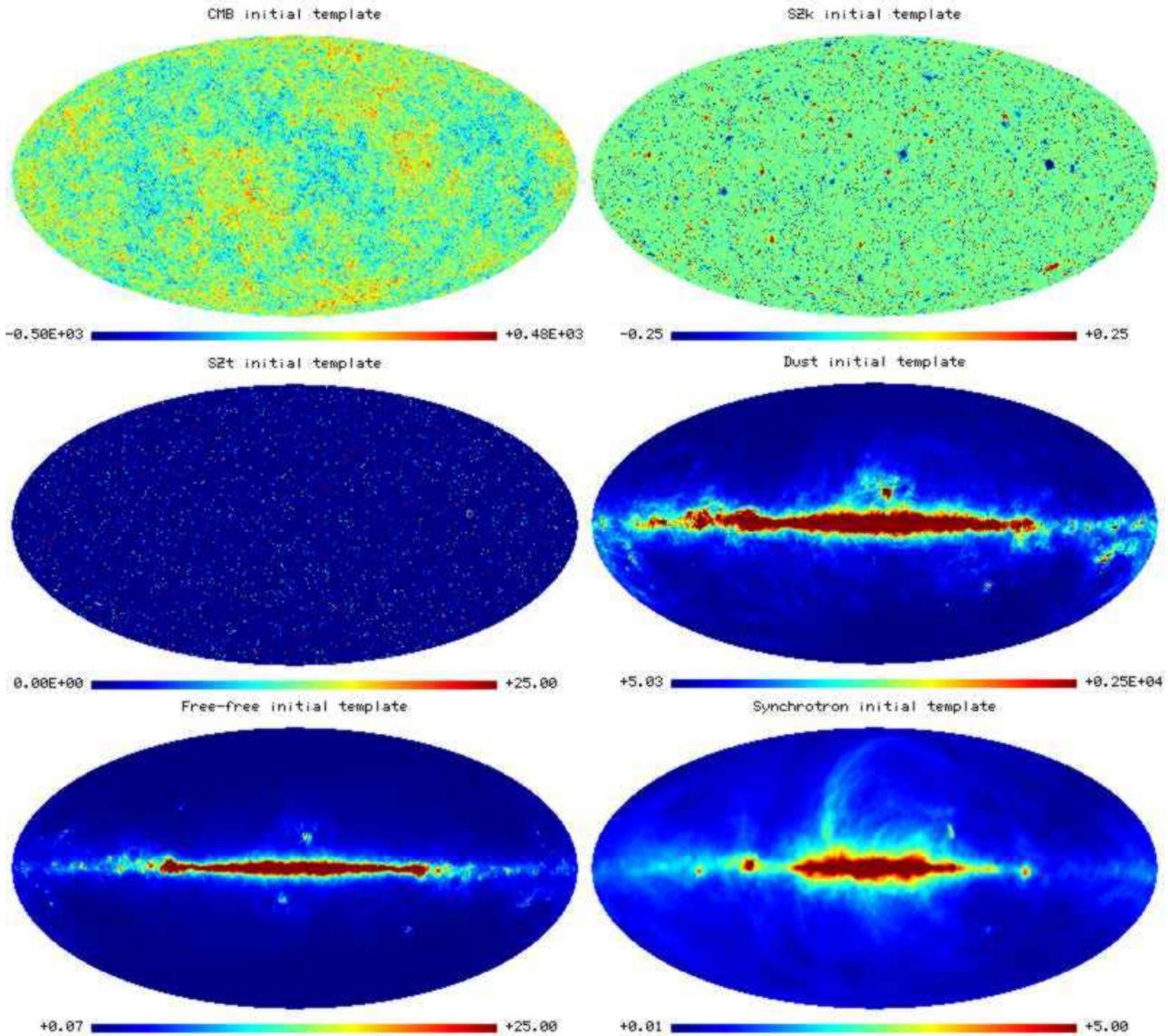,
width=17cm}}
%\centerline{
%\epsfig{file=CMB_init.ps,
%angle=-90,width=8.5cm}
%\quad 
%\epsfig{file=SZk_init.ps,
%angle=-90,width=8.5cm}
%}
%\centerline{
%\epsfig{file=SZt_init.ps,
%angle=-90,width=8.5cm}
%\quad 
%\epsfig{file=Dust_init.ps,
%angle=-90,width=8.5cm}
%}
%\centerline{
%\epsfig{file=FF_init.ps,
%angle=-90,width=8.5cm}
%\quad 
%\epsfig{file=Synch_init.ps,
%angle=-90,width=8.5cm}
%}
\caption{All-sky realisations of the six main physical components
contributing to the sky emission at Planck wavelengths.
The components are primordial CMB, kinetic and thermal SZ effects
from clusters and Galactic dust, free-free and synchrotron emission. 
Each map is defined in the {\sc HEALPix} pixelisation scheme with
$N_{\rm side} = 2048$, which corresponds to $\sim 50 \times 10^6$
pixels of size 1.7 arcmin. Each map is plotted at 300 GHz 
in units of $\mu$K.}
\label{fig1}
\end{center}
\end{figure*}

Each of the six physical components of emission are simulated in the
{\sc HEALPix}\footnote{http://www.eso.org/kgorski/healpix/}
pixelisation scheme (Gorski, Hivon \& Wandelt 1999) with
$N_{\rm side} = 2048$, which corresponds to $\sim$ 50 million pixels
of size 1.7 arcmin, or $\ell_{\rm max} = 3N_{\rm side}-1 = 6143$.  
Since the highest
angular resolution of the Planck satellite is 4.5 arcmin, a pixel size
of 1.7 arcmin ensures that the fields are Nyquist sampled. The models
used for each physical component are discussed below. These input maps
are plotted in Fig.~\ref{fig1} at 300 GHz, and 
the corresponding power spectra are shown in Fig.~\ref{fig2}.

\subsection{CMB anisotropy}

The primordial CMB anisotropy field is generated over the whole sky
using the {\sc Synfast} routine in the {\sc  HEALPix} software package, and
corresponds a Gaussian realisation of a spatially-flat standard
inflationary CDM model with $\Omega_{\rm m} = 0.35$, $\Omega_\Lambda
= 0.65$, $\Omega_{\rm b} h^2 = 0.02$ and $h=0.65$,
for which the $C_\ell$ coefficients were created using {\sc Cmbfast}
(Seljak \& Zaldarriaga 1996).

\subsection{Thermal and kinetic SZ effects}

The thermal SZ effects from individual clusters were simulated using
the gas dynamics code of Eke, Navarro \& Frenk (1998), from which 210 cluster
templates were obtained. These templates were distributed on the sky
according to a Poisson distribution and with random orientations. The
redshift distribution was chosen to be consistent with the
Press-Schechter model (Press \& Schechter 1974) for a $\Omega_{\rm m}
= 0.35$, $\Omega_\Lambda=0.65$ cosmology (R. Kneissl, private communication). 
To simulate the kinetic SZ
effect, the cluster radial velocities are assumed to be Gaussian
distributed with a dispersion of 400 km s$^{-1}$ at $z=0$.

\subsection{Galactic dust emission}
\label{Dust}

The Galactic dust contribution is modelled using the DIRBE-IRAS
100-$\mu$m dust map (Schlegel, Finkbeiner \& Davis 1998).  The angular
resolution of the map is around 5 arcmin, which is comparable with
Planck resolution in the highest frequency channels. 
The emission at each Planck frequency can be predicted by extrapolating
the 100-$\mu$m flux assuming a one-component dust model
with a temperature $T_{\rm dust}$ = 18 K and dust emissivity $\beta =
2$. The colour correction factor for the DIRBE 100-$\mu$m filter
was also taken into account; (see Finkbeiner, Davis \& Schlegel
1999).

\subsection{Galactic synchrotron emission}

The basic template of the synchrotron emission is the destriped
version of the 408 MHz Haslam survey (Haslam et al.  1982), to which
additional structure was artificially added at sub-degree angular
scales by extrapolating the angular power spectrum of the Haslam
survey as $C_\ell \propto \ell^{-3}$.  The synchrotron map at 300
GHz is obtained by extrapolating the 408 MHz survey using an all-sky
map of the spectral index constructed by combining the low frequency
surveys at 408 MHz, 1420 MHz (Reich \& Reich 1986) and 2326 MHz
(Jonas, Baart \& Nicholson 1998) and 
padding the unobserved area around the South pole
with the mean spectral index, at a resolution (FWHM) of 10 degrees
(G. Giardino, private communication).  To predict the synchrotron
contribution at the Planck observing frequencies, a constant intensity
spectral index $\beta = -0.9$ was assumed.

\subsection{Galactic free-free emission}

Reliable maps of Galactic free-free emission are not currently
available, although experiments such as the H$\alpha$ Sky
Survey\footnote{ http://www.swarthmore.edu/Home/News/Astronomy/} 
(Gaustad et al. 2001) and the Wisconsin H$\alpha$ Mapper --- WHAM 
\footnote{http://www.astro.wisc.edu/wham/} (Haffner 2001) should soon provide
maps that could be used as templates. The resolution of the surveys is
about 0.8 arcmin for H$\alpha$ Sky Survey and about 1 degree for WHAM.
For the time being, however, we
create a free-free template based on the DIRBE/IRAS dust map, of which
60 per cent of the emission is a dust-correlated component and 40 per
cent is uncorrelated. The uncorrelated component is simply a dust map
flipped north-south in Galactic coordinates. The spectral index of the
free-free emission is assumed to be $\beta = -0.16$ and the
normalisation is that suggested by Bouchet \& Gispert (1999).

\begin{figure}
\begin{center}
\centerline{
\epsfig{file=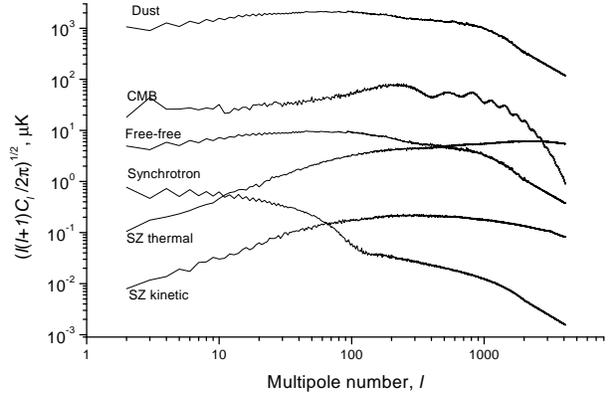,
width=9cm}}
\caption{Power spectra of the input maps shown in Fig.~\ref{fig1}}
\label{fig2}
\end{center}
\end{figure}

\section{Simulated Planck observations} 
\label{simobs}

\begin{figure}
\begin{center}
\centerline{
\epsfig{file=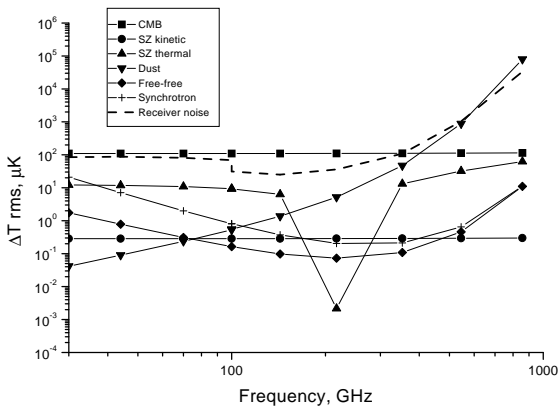,
width=9cm}}
\caption{The rms equivalent thermodynamic temperature fluctuation per pixel at
each Planck observing frequency in the region of sky with
Galactic latitude $65^o < b < 75^o$.  
The rms instrumental noise per pixel in each frequency channel is also plotted.
The pixel size is 1.7 arcmin.}
\label{fig3}
\end{center}
\end{figure}

If we observe the sky at $n_f$ observing frequencies then, in any
given direction $\hvect{x}$, we obtain a {\em data vector} of length
$n_f$ that contains the observed temperature fluctuation in this
direction at each observing frequency plus instrumental noise.  In
order to relate this data vector to the emission from each physical
component it is useful to introduce the $n_f\times n_c$ {\em frequency
response matrix} with components defined by
\[
F_{\nu p} = \int_0^\infty t_\nu(\nu') f_p(\nu') \,{\rm d}\nu',
\]
where $t_\nu(\nu')$ is the frequency response (or transmission) of the
$\nu$th frequency channel. Assuming that the satellite observing beam in each
channel is spatially-invariant, we may write the beam-smoothing
as a convolution and the $\nu$th component of
the data vector in the direction $\hvect{x}$ is then given by
\begin{equation}
d_\nu(\hvect{x}) = \int_{4\pi} B_\nu(\hvect{x}\cdot\hvect{x}') 
\sum_{p=1}^{n_c} F_{\nu p}\,s_p(\hvect{x}') \,{\rm d}\Omega'
+ \epsilon_\nu(\hvect{x})
\label{datadef}
\end{equation}
where $B_\nu$ is the beam profile for the $\nu$th frequency
channel. The beam profile at each frequency
is assumed to be Gaussian. The
$\epsilon_\nu(\hvect{x})$ term represents the instrumental noise on
the observations in
the $\nu$th channel, which for simplicity
we assume to be uncorrelated Gaussian noise with a fixed rms over
the whole sky. The 
assumed observational parameters of the Planck satellite are
given in Table~\ref{table1}.
\begin{table*}
\begin{center}
\caption{The assumed observational parameters for the Planck Surveyor
satellite. Angular resolution is quoted as FWHM 
for a Gaussian beam. Sensitivities are quoted per
beam FWHM for 12 months of observation.}
\label{table1}
\begin{tabular}{lccccccccccc} \hline
& \multicolumn{4}{c}{Low Frequency Instrument} 
& & \multicolumn{6}{c}{High Frequency Instrument} \\ \hline
Central frequency (GHz):   
%& 31.5 & 53   & 90   & 125  & 143  & 217  & 353  & 545  & 857 \\
& 30   & 44   & 70   & 100  & & 100  & 143  & 217  & 353  & 545  & 857 \\
Fractional bandwidth ($\Delta\nu/\nu$):   
%& 0.15 & 0.15 & 0.15 & 0.15 & 0.37 & 0.37 & 0.37 & 0.37 & 0.37 \\
& 0.2 & 0.2 & 0.2 & 0.2 & & 0.37 & 0.37 & 0.37 & 0.37 & 0.37 & 0.37 \\
Transmission:
%& 1.0  & 1.0  & 1.0  & 1.0  & 0.3  & 0.3  & 0.3  & 0.3  & 0.3 \\
& 1.0  & 1.0  & 1.0  & 1.0  & & 0.3  & 0.3  & 0.3  & 0.3  & 0.3  & 0.3 \\
Angular resolution (arcmin):  
%& 30   & 18   & 12   & 12   & 10.3 & 7.1  & 4.4  & 4.4  & 4.4 \\
& 33   & 23   & 14   & 10   & & 10.6   & 7.4  & 4.9  & 4.5  & 4.5  & 4.5 \\
$\Delta T$ sensitivity ($\mu$K): 
%& 21.2 & 20.4 & 39.2 & 96.5 & 3.3  & 5.5  & 32.9 & 209  & 11356 \\
& 4.4  & 6.5  & 9.8 & 11.7 & & 4.9 & 5.7 & 12.5  & 40.9 & 392  & 12621 \\
\hline
\end{tabular}
\end{center}
\end{table*}
Fig.~\ref{fig3} shows the rms equivalent thermodynamic temperature
fluctuation at each Planck frequency due to each physical component,
in the sky region lying between Galactic latitudes $65^o< b < 75^o$.
The rms instrumental noise per pixel in each frequency channel is also plotted.

%

%\section{Statistical properties of the data}

If the beam profile at each frequency is spatially-invariant and
circularly-symmetric, it is convenient to work in terms of spherical
harmonic coefficients. We now consider the statistical properties of
these coefficients.

\subsection{All-sky observations}

If each data map $d_\nu(\hvect{x})$ is defined over the whole sky,
then on performing a spherical harmonic transform the convolution in
(\ref{datadef}) becomes a simple multiplication. Thus, we obtain
\begin{equation}
d^{(\nu)}_{\ell m} = \sum_{p=1}^{n_c} R^{(\nu p)}_{\ell}a^{(p)}_{\ell m}
+\epsilon^{(\nu)}_{\ell m},
\label{dataft}
\end{equation}
where we have adopted the usual notation for spherical harmonic
coefficients $f_{\ell m} = \int_{4\pi} {\rm d}\Omega\, Y^*_{\ell
m}(\hvect{x})f(\hvect{x})$ in which $Y_{\ell m}(\hvect{x})$ is a 
standard spherical harmonic function.  
Thus, $d^{(\nu)}_{\ell m}$ are the spherical
harmonic coefficients of the $\nu$th frequency map. These are related
to the spherical harmonic coefficients $a^{(p)}_{\ell m}$ of the $p$th
physical component via the {\em response matrix} for the observations
$R^{(\nu p)}_\ell = B^{(\nu)}_\ell F_{\nu p}$, where $B^{(\nu)}_\ell$ are the
harmonic coefficients of the $\nu$th observing beam. Finally,
$\epsilon^{(\nu)}_{\ell m}$ is the instrumental noise on the harmonic mode
$(\ell,m)$.  It is important to note that (\ref{dataft}) is satisfied
at each spherical harmonic mode $(\ell,m)$ {\em independently}.  Thus,
in matrix notation, at each mode we have
\begin{equation}
{\mathbfss d}_{\ell m} = {\mathbfss R}_\ell {\mathbfss a}_{\ell m}
+\bmath{\epsilon}_{\ell m},
\label{dataft2}
\end{equation}
where ${\mathbfss d}_{\ell m}$, ${\mathbfss a}_{\ell m}$ and
$\bmath{\epsilon}_{\ell m}$ are column vectors containing $n_f$, $n_c$
and $n_f$ complex components respectively, and the response matrix
${\mathbfss R}_\ell$ has dimensions $n_f\times n_c$.

Although the data at a given $(\ell,m)$ mode do not depend on the
harmonics of the physical components at other modes, the a priori
covariance structures of the harmonic coefficients of the components
$a^{(p)}_{\ell m}$ and the instrumental noise $\epsilon^{(\nu)}_{\ell m}$ do
correlate different modes.  The sky emission (from the Galactic
components) is anisotropic owing to the presence of pronounced
emission from the Galactic plane. Furthermore, scanning strategies
with non-uniform coverage of the sky lead to spatial variations in the
noise rms per pixel, and so the instrumental noise field is also
anisotropic.  This results in a priori correlations between different
$(\ell,m)$ modes, so that in general we have
\begin{eqnarray}
\langle {\mathbfss a}_{\ell m} {\mathbfss a}^\dagger_{\ell' m'}
\rangle & = & {\mathbfss C}_{\ell m,\ell'm'}, \label{cmat} \\ 
\langle \bmath{\epsilon}_{\ell m}\bmath{\epsilon}^\dagger_{\ell' m'} \rangle
& = & {\mathbfss N}_{\ell m,\ell'm'}, \label{nmat}
\end{eqnarray}
where we have defined the set of ensemble-average signal and noise
covariance matrices ${\mathbfss C}_{\ell m,\ell'm'}$ and ${\mathbfss
N}_{\ell m,\ell'm'}$. Each matrix ${\mathbfss C}_{\ell m,\ell'm'}$ has
dimensions $n_c\times n_c$ and contains the a priori covariances of
the harmonic coefficients of the different physical components at the
modes $(\ell,m)$ and $(\ell',m')$. Each matrix ${\mathbfss N}_{\ell
m,\ell'm'}$ has dimensions $n_f \times n_f$ and contains the
covariances between the modes of the instrumental noise harmonics at
the different observing frequencies.  It should also be remembered
that the emission from components other than the CMB is highly
non-Gaussian and so these second-order statistics will not, in
general, provide a full statistical description of the sky emission,
particularly in the Galactic plane. In principle, we can also
include any prior information about noise correlation properties
in the noise covariance
matrix $\mathbfss N$. A typical example is non-homogeneous pixel noise in each
frequency channel.

\subsection{Galactic cuts}

Since Planck is primarily a CMB instrument, `component separation' is
often taken  to mean `foreground removal'. In particular, mapping of
Galactic emission is regarded only as a secondary goal of the mission.
When performing an all-sky foreground removal, one might therefore consider
first removing the pronounced emission from the Galactic plane by
imposing a Galactic cut on each individual frequency map. The new
input data maps are then given by
\begin{equation}
\tilde{d}_\nu(\hvect{x}) = W(\hvect{x})d_\nu(\hvect{x}),
\label{windowdef}
\end{equation}
where the window $W(\hvect{x})$ is zero in the Galactic cut and equals
unity outside it (although more general windows are possible).  As
indicated in (\ref{windowdef}), the window $W$ is usually chosen to be the same
for each observing frequency, and is often taken to be symmetric about
the Galactic equator (for ease of computation, as discussed
below). Indeed, this approach was adopted by Prunet et al. (2001).

The imposition of a Galactic cut does, however, produce some
troublesome effects. Firstly, the spherical harmonic basis $Y_{\ell
m}(\hvect{x})$ is no longer orthonormal over the cut-sphere. For the
Wiener filter approach, in particular, this leads to inconsistencies
between performing the component separation in real space and harmonic
space (see Prunet et al.). Secondly, the Galactic cut leads to
additional correlations between harmonic coefficients at different
$(\ell,m)$ modes. For a function $f(\hvect{x})$ defined on the sphere,
the harmonic modes after the Galactic cut $\tilde{f}=Wf$ are given by
\[
\tilde{f}_{\ell m} = \int_{4\pi} {\rm d}\Omega\, Y^*_{\ell m}(\hvect{x})
W(\hvect{x}) f(\hvect{x}) = \sum_{\ell',m'} W_{\ell m,\ell' m'}
f_{\ell' m'},
\]
where the coupling matrix is
\[
W_{\ell m,\ell' m'} = \int_{4\pi} {\rm d}\Omega\, W(\hvect{x})Y^*_{\ell
m}(\hvect{x}) Y_{\ell' m'}(\hvect{x}).
\]

The problem of loss of orthonormality is usually solved by performing
the entire analysis in terms of a new basis $Y'_{\ell m}(\hvect{x})$
that is orthonormal on the cut-sky, such that
\[
\int_{4\pi} {\rm d}\Omega\, W(\hvect{x}){Y'}^*_{\ell m}(\hvect{x}) Y'_{\ell'
m'}(\hvect{x}) = \delta_{\ell\ell'}\delta_{mm'}.
\]
For $\ell_{\rm max} \la 50$, this basis can be constructed by
performing a Cholesky decomposition of the coupling matrix (Gorski
1994). For higher $\ell_{\rm max}$, a singular value decomposition
(SVD) is required (Tegmark 1997; 
Mortlock, Challinor \& Hobson 2001). We note
that, for simulated MAP data with $\ell_{\rm max} = 512$,
Prunet et al.  find the computational and memory costs of computing
the new orthonormal basis to be reasonable, but the corresponding
computing requirements for Planck will be somewhat larger.
Perhaps more importantly, the computation of the new orthonormal basis
is entirely unfeasible for either MAP or Planck unless the edges of
the Galactic cut are chosen to be lines of constant latitude. It
should also be remembered that for the new basis functions the
subscripts $\ell$ and $m$ no longer have their usual physical
meanings. Nevertheless, for large $\ell_{\rm max}$ (such as for MAP or
Planck) the new basis functions $Y'_{\ell m}(\hvect{x})$ are very
close to zero in the cut, and so the coefficients of these modes have
the desirable property of being insensitive to emission from the cut
region.

Unfortunately, the construction of the new orthonormal basis does not
solve the problem of mode coupling.  The 
new basis can be expressed as a linear combination of
the standard spherical harmonic basis functions,
\[
Y'_{\ell m} (\hvect{x}) = \sum_{\ell',m'} O_{\ell m,\ell' m'} Y_{\ell'
m'}(\hvect{x}),
\]
where matrix elements $O_{\ell m,\ell' m'}$ are given by the `overlap'
integrals of the new and old basis functions,
\[
O_{\ell m,\ell' m'} = \int_{4\pi} {\rm d}\Omega\, {Y'}^*_{\ell m}(\hvect{x})
Y_{\ell' m'}(\hvect{x}).
\]
If we expand the sky emission from the physical components in terms of
our new basis, we thus obtain
\begin{equation}
{\mathbfss a}'_{\ell m} = \sum_{\ell',m'} O_{\ell m,\ell' m'}
{\mathbfss a}_{\ell' m'},
\label{aprime}
\end{equation}
and a similar expression holds for the coefficients
$\bmath{\epsilon}'_{\ell m}$ of the instrumental noise field at each
observing frequency.  Hence, the a priori correlation structure in the
new basis is given by
\begin{eqnarray}
\langle {\mathbfss a}'_{\ell m} {{\mathbfss a}'}^\dagger_{\ell' m'}
\rangle & = & {\mathbfss C}'_{\ell m,\ell'm'}, \label{cprime} \\ 
\langle \bmath{\epsilon}'_{\ell m} {\bmath{\epsilon}'}^\dagger_{\ell' m'}
\rangle & = & {\mathbfss N}'_{\ell m,\ell'm'}, \label{nprime}
\end{eqnarray}
where explicit expressions for ${\mathbfss C}'_{\ell m,\ell'm'}$ and
${\mathbfss N}'_{\ell m,\ell'm'}$ are easily obtained from (\ref{aprime}),
(\ref{cmat}) and (\ref{nmat}). 
In general, coupling still persists betweem different
$(\ell,m)$ modes.

An additional problematic feature of performing a Galactic cut is that
the operation of beam convolution is no longer diagonal in the new
orthonormal basis.  From (\ref{dataft}) and (\ref{aprime}), we have
\begin{equation}
{d'}^{(\nu)}_{\ell m} = \sum_{p=1}^{n_c} \sum_{\ell' m'}
R^{(\nu p)}_{\ell'} O_{\ell m,\ell' m'} \,a^{(p)}_{\ell' m'} 
+{\epsilon'}^{(\nu)}_{\ell m},
\label{dprime}
\end{equation}
which, in general, {\em cannot} be written in the form
\begin{equation}
{\mathbfss d}'_{\ell m} = {\mathbfss R}'_{\ell m} {\mathbfss a}'_{\ell m}
+\bmath{\epsilon}'_{\ell m}.
\label{dprime2}
\end{equation}
and this leads to further coupling between different $(\ell,m)$ modes.
Thus, in general, the operation of beam convolution is not merely
non-diagonal, but is not strictly defined in a formal sense.
We note, however, that we can write (\ref{dprime}) in this form in the
special case where the response matrix ${\mathbfss R}$ is not a
function of $\ell$. This corresponds to no beam convolution.

\subsection{Incomplete sky coverage}

Similar considerations to the above also apply to incomplete sky
coverage resulting from, for example, the scanning strategy of the
satellite or instrument failure. In this case, however, $W$ can be
highly irregular in shape and different for each observing
frequency. If the affected regions are not too large one could define
a single window $W$ that encompasses the unobserved regions at all
frequencies, but this would probably not be viable in
practice. Moreover, it is unlikely that one could find an adequate
Galactic cut with edges at constant latitude, and so it would be
computationally impossible to construct an new orthonormal basis on
the cut-sky.

\subsection{Choice of basis functions}

Following the above discussion, we have the choice of performing the
calculation in terms of the standard spherical harmonic basis
functions $Y_{\ell m}$ or some other basis $Y'_{\ell m}$ that is
orthonormal on the part of the sky outside some constant latitude
Galactic cut. In both bases, there is coupling of modes through their
a priori covariance structure, as defined by (\ref{cmat}--\ref{nmat}) 
or (\ref{cprime}--\ref{nprime}) respectively.
Thus, in either case, a general analysis of the data cannot be
performed `mode-by-mode', but requires one to estimate the full
`signal' vector ${\mathbfss a}$ (or ${\mathbfss a}'$) simultaneously
from the full data vector ${\mathbfss d}$ (or ${\mathbfss d}'$); these
vectors can be regarded simply as the concatenation of the vectors
${\mathbfss a}_{\ell m}$ and ${\mathbfss d}_{\ell m}$ respectively for
all possible values of $\ell$ and $m$ (or their primed counterparts).

It is also worth noting, however, that in the cut-sky basis
one must {\em assume} that (\ref{dprime2}) holds, which is
not valid unless there is no beam convolution. This
assumption is, of course, unnecessary in the standard harmonic basis,
since (\ref{dataft2}) holds exactly for all-sky observations. The assumption
(\ref{dprime2}) will, in general, lead to reconstruction artefacts which are
most pronounced near the edges of the Galactic cut.

Owing to the above difficulties of working in the cut-sky basis, and
given that we are interested in recovering maps of the Galactic
components as well as the CMB, in this paper we choose {\em not} to
impose a Galactic cut, and instead attempt to reconstruct the physical
components over the {\em whole sky}. Thus we perform our calculations
in the standard spherical harmonic basis.

\section{Component separation}
\label{compsep}

The aim of any component separation algorithm is to use the data maps
$d_\nu(\hvect{x})$ at each frequency to obtain an estimated maps
$\hat{s}_p(\hvect{x})$ of the emission from each physical component.
Typical methods include singular-valued decomposition, Wiener
filtering or the maximum-entropy method, all of which can be viewed in
Bayesian context (see Paper I).

\subsection{Harmonic-space maximum-entropy method}

A complete description of the `mode-by-mode' Fourier-space MEM
component separation algorithm as applied to small patches of sky is
given in Paper I. The main difference in applying the technique to
all-sky reconstructions is that Fourier transforms must be replaced by
spherical harmonic transforms.  Aside from this modification, the
algorithm remains basically unchanged, and so we give only a brief
outline here.

Let us consider the standard (unprimed) spherical harmonic
basis, although the following discussion is equally valid for the
primed basis described above. Using
 Bayes' theorem, the estimator $\hat{\mathbfss a}$
of the signal vector is usually taken to be that which maximises the
posterior probability
\[
\Pr({\mathbfss a}|{\mathbfss d}) \propto \Pr({\mathbfss d}|{\mathbfss
a})\Pr({\mathbfss a})
\]
where $\Pr({\mathbfss d}|{\mathbfss a})$ is the likelihood of
obtaining the data given a particular signal vector and
$\Pr({\mathbfss a})$ is the prior probability that codifies our
expectations about the signal vector before acquiring any data.
Indeed, it is the specification of the prior alone that differentiates
between different component separation algorithms such as
singular-value decomposition, Wiener filtering and the maximum-entropy
method (see Paper I).

Unfortunately, the 
lengths of the (complex) data and signal vectors are $\sim n_f
l_{\rm max}^2$ and $\sim n_c l_{\rm max}^2$, and so for MAP or Planck
observations the dimensionality of the numerical maximisation problem
renders it computationally unfeasible. It is therefore necessary to
make some approximation.
The most straightforward approximation is to neglect any coupling
between different $(\ell,m)$ modes, and perform the reconstruction
independently `mode-by-mode'.  Although this may appear an extreme
approximation, we will see in Section~\ref{apply} that it still produces
excellent reconstructions. Taking the modes to be independent
corresponds to assuming that the
likelihood and prior probability distributions factorise, such that
\begin{eqnarray}
\Pr({\mathbfss d}|{\mathbfss a}) & = & \prod_{\ell, m} \Pr({\mathbfss
d}_{\ell m}|{\mathbfss a}_{\ell m}), \label{likeprod}\\ 
\Pr({\mathbfss a}) & = &
\prod_{\ell, m} \Pr({\mathbfss a}_{\ell m}). \label{priorprod}
\end{eqnarray}
This offers an enormous computational advantage, since one can
maximise the posterior probability
\[
\Pr({\mathbfss a}_{\ell m}|{\mathbfss d}_{\ell m}) \propto
\Pr({\mathbfss d}_{\ell m}|{\mathbfss a}_{\ell m}) \Pr({\mathbfss
a}_{\ell m})
\]
at each mode separately. Thus one replaces a single numerical
maximisation over a parameter space with $2n_c l_{\rm max}^2$ real
dimensions by $l_{\rm max}^2$ {\em separate} minimsations with $2n_c$
real dimensions.

The factorisations (\ref{likeprod}) and (\ref{priorprod}) imply 
an assumed a priori covariance
structure of the form
\begin{eqnarray*}
\langle {\mathbfss a}_{\ell m} {\mathbfss a}^\dagger_{\ell' m'} \rangle
& = & {\mathbfss C}_{\ell m} \delta_{\ell\ell'}\delta_{mm'}, \\ 
\langle \bmath{\epsilon}_{\ell m} \bmath{\epsilon}^\dagger_{\ell' m'} \rangle &
= & {\mathbfss N}_{\ell m} \delta_{\ell\ell'}\delta_{mm'}.
\end{eqnarray*}
Although not required by the mathematical formalism, we have made the
additional simplifying assumption that ${\mathbfss C}_{\ell m}$
and ${\mathbfss N}_{\ell m}$ do not depend on $m$, so that
\begin{eqnarray}
\langle {\mathbfss a}_{\ell m} {\mathbfss a}^\dagger_{\ell' m'} \rangle
& = & {\mathbfss C}_\ell \delta_{\ell\ell'}\delta_{mm'}, \label{csimp}\\ 
\langle \bmath{\epsilon}_{\ell m} \bmath{\epsilon}^\dagger_{\ell' m'} \rangle &
= & {\mathbfss N}_\ell \delta_{\ell\ell'}\delta_{mm'}.\label{nsimp}
\end{eqnarray}
In the standard spherical harmonic basis, this approximation is 
equivalent to assuming
that the emission from each physical component and the instrumental
noise are all isotropic random fields on the celestial sphere. The
matrices ${\mathbfss C}_\ell$ and ${\mathbfss N}_\ell$ contain the
(cross) power spectra of, respectively, the physical components at
$\nu_0$ and the instrumental noise at each observing frequency. 
Unfortunately, the analogous approximation in the cut-sky basis does
not have such a straightforward interpretation.

At each $(\ell, m)$ mode, any a priori correlation information is
contained in the matrix ${\mathbfss C}_\ell$.  As explained in Paper
I, this information is most easily incorporated into the
reconstruction algorithm by first defining a `hidden' vector
${\mathbfss h}_{\ell m}$ at each mode, that is related to the vector
${\mathbfss a}_{\ell m}$ by
\begin{equation}
{\mathbfss a}_{\ell m} = {\mathbfss L}_\ell {\mathbfss h}_{\ell m}.
\label{hidden}
\end{equation}
The $n_c \times n_c$ lower triangular matrix ${\mathbfss L}_\ell$ is
obtained by performing the Cholesky decomposition ${\mathbfss C}_\ell
= {\mathbfss L}_\ell {\mathbfss L}^\dagger_\ell$.  The components of
the hidden vector thus have the useful property of being a priori
uncorrelated with unit variance, since then
\[
\langle {\mathbfss a}_{\ell m}{\mathbfss a}_{\ell m}^\dagger\rangle
= \langle {\mathbfss L}_\ell{\mathbfss h}_{\ell m}
{\mathbfss h}_{\ell m}^\dagger 
{\mathbfss L}_\ell^\dagger\rangle = {\mathbfss
L}_\ell\langle{\mathbfss h}_{\ell m}{\mathbfss h}_{\ell m}^\dagger\rangle 
{\mathbfss L}_\ell^\dagger = {\mathbfss C}_\ell.
\]
The reconstruction is then performed entirely in terms of
${\mathbfss h}_{\ell m}$ and the corresponding reconstructed signal
vector is subsequently found using (\ref{hidden}).

Assuming the instrumental noise to be Gaussian, the likelihood
function at each mode is given by
\[
\Pr({\mathbfss d}_{\ell m}|{\mathbfss h}_{\ell m}) \propto \exp
\left[-\chi^2({\mathbfss h}_{\ell m})\right],
\]
where $\chi^2$ is the standard misfit statistic given by
\[
\chi^2({\mathbfss h}_{\ell m}) = ({\mathbfss d}_{\ell m}-{\mathbfss
R}_\ell {\mathbfss L}_\ell {\mathbfss h}_{\ell m})^\dag {\mathbfss
N}_\ell^{-1} ({\mathbfss d}_{\ell m}-{\mathbfss R}_\ell {\mathbfss
L}_\ell {\mathbfss h}_{\ell m})
\]
For the prior $\Pr({\mathbfss h}_{\ell m})$, we
assume the entropic form
\[
\Pr({\mathbfss h}_{\ell m}) \propto \exp[\alpha S({\mathbfss h}_{\ell
m},{\mathbfss m})]
\]
where $S({\mathbfss h}_{\ell m},{\mathbfss m})$ is the cross entropy
of the complex vectors ${\mathbfss h}_{\ell m}$ and ${\mathbfss m}$,
where ${\mathbfss m}$ is a model vector to which ${\mathbfss h}_{\ell
m}$ defaults in absence of data. The form of the cross entropy for
complex images is given in Paper I. Thus, maximising the posterior
probability $\Pr({\mathbfss h}_{\ell m}|{\mathbfss d}_{\ell m})$ with
respect to ${\mathbfss h}_{\ell m}$ is equivalent to minimising the
function
\begin{equation}
\Phi({\mathbfss h}_{\ell m})=\chi^2({\mathbfss h}_{\ell m}) - \alpha
S({\mathbfss h}_{\ell m},{\mathbfss m}).
\label{function}
\end{equation}

As explained in Paper I, the regularising parameter $\alpha$ can be
determined in a Bayesian manner by treating it simply as another
parameter in the hypothesis space. Indeed, by assuming a Gaussian form
for the posterior probability distribution near its maximum, one can
derive a closed-form expression for the optimal value of $\alpha$ (see
Paper I). Since the calculation is performed at each $(\ell,m)$ mode
independently, one could in principle obtain the optimal value of
$\alpha$ separately for each mode. However, since the dimensionality
of the parameter space at each mode is only $2n_c$-dimensional
(i.e. 12-dimensional in our case), the Gaussian approximation to the
posterior distribution is occasionally too inaccurate to determine
$\alpha$ properly. Nevertheless, we would not expect the optimal level
of regularisation to depend strongly on $m$, but to vary considerably with
$\ell$. Thus we assume the optimal value of $\alpha$ to be the same
for all modes at the same $\ell$. Taking these modes together, the
dimensionality of the parameter space it then sufficiently large that
the Gaussian approximation to the posterior distribution is reasonably
accurate, and the optimal value of $\alpha$ is well-determined. Thus,
the harmonic-space MEM algorithm employs optimal, scale-dependent
regularisation, which allows for a more accurate reconstruction.

The numerical minimisation of (\ref{function}) can be performed using a
variable metric minimiser (Press et al. 1994). As a result of the
independence of each $(\ell, m)$ mode, parallelisation of the
separation algorithm is straightforward.  Full-resolution Planck
simulations in {\sc HEALPix} with $N_{\rm side}=2048$ have a
maximum multipole $\ell_{\rm max}=3N_{\rm side} - 1 =  6143$.
The entire component separation can be  performed in $\sim$ 6
hr using 30 R10000 processors on the {\sc `Cosmos'} SGI Origin 2000 
supercomputer and required only 14 Gb of RAM.

As shown in Paper I, one can also estimate the covariance matrix of
the reconstruction errors on the harmonic modes ${\mathbfss a}_{\ell
m}$. Making a Gaussian approximation to the posterior distribution at
each mode, we find
\begin{equation}
\langle (\hat{\mathbfss a}_{\ell m}-{\mathbfss a}_{\ell m})
(\hat{\mathbfss a}_{\ell m}-{\mathbfss a}_{\ell m})^\dagger \rangle =
{\mathbfss L}_\ell {\mathbfss H}^{-1}_{\ell m} {\mathbfss
L}^\dagger_\ell,
\label{hesserr}
\end{equation}
where 
${\mathbfss H}_{\ell m} = \nabla\nabla\Phi({\mathbfss h}_{\ell m})$ 
is the Hessian matrix of the posterior
distribution at its peak $\hat{\mathbfss h}_{\ell m}$.  For
full-resolution Planck data, the calculation of the errors on the
component separation required just 15 mins additional computation
time on {\sc Cosmos}.

\section{Application to simulated data}
\label{apply}

We now apply the separation algorithm outlined above to the simulated
Planck Surveyor observations 
described in Section \ref{simobs}. Owing to the limited resolution of
our foreground maps, we only consider multipoles up to
$\ell_{\rm max}=4096$, which is the Nyquist sampling limit for
pixels of size 1.7 arcmin. This reduces both the computation time 
and memory required
by about a factor of two, as compared with the 
case $\ell_{\rm max}=6143$.

Clearly, the quality of
the reconstructions will depend on our prior knowledge of the input
components. Following Paper I, we assume that the frequency spectrum
behaviour of the components is accurately known. 
Nevertheless, as mentioned in the
introduction, the relaxation of this assumption has investigated by
Jones et al. (1999), who showed that it is possible to accommodate
mild uncertainties in frequency behaviour and spatial variation of 
spectral indices. Prior power
information may be incorporated into the algorithm through the signal
covariance matrix ${\mathbfss C}_\ell$ given in
(\ref{csimp}). 
Similarly, any knowledge of the correlation structure of the
instrumental noise can be included via the noise covariance matrix
${\mathbfss N}_\ell$. In our simulated data, however, we assume
homogeneous, uncorrelated instrumental noise in each frequency
channel.

\subsection{Partial prior covariance information}

As an illustration, we begin by assuming knowledge of the
azimuthally-averaged power spectra of each input components, as
shown in Fig.~\ref{fig2}, {\em except} that of the CMB. Since the
accurate determination of the positions
and heights of the acoustic peaks in the CMB spectrum is of such
fundamental importance, we choose not to prejudice our reconstruction
by supplying the component separation algorithm with the 
power spectrum of the input CMB map. Thus, as our assumed CMB power
spectrum, we take a simple functional form which matches the overall
shape of the true spectrum, but contains no acoustic peaks (this
assumed power spectrum is plotted as the dotted line in the
first panel of Fig.\ref{fig8}).

The inclusion of the azimuthally-averaged input power spectra for the
remaining components is reasonable, and would be possible in
practice. As discussed in section \ref{model}, the maps of the kinetic and
thermal SZ effect have been produced with some care, and we would
expect them to provide a reasonably accurate representation of the
true power spectra for these two components. Alternatively, one could
use an average power spectrum for each SZ component obtained from a
large number of simulated realisations, which would be statistically
more accurate.  For the Galactic components, the dust and synchrotron
maps are simply taken from existing observations (albeit with some
added structure on small scales). Unfortunately, considerable uncertainty 
remains for Galactic emission in the free-free. Nevertheless, 
free-free observations should be available
before the Planck data are analysed, so that an approximate free-free
power spectrum could be calculated.  As discussed in section~\ref{model},
for the time being, we create a free-free template based on the
DIRBE/IRAS dust map, of which 60 per cent of the emission is a
dust-correlated component and 40 per cent is uncorrelated, with the
uncorrelated component being simply the dust map flipped north-south
in Galactic coordinates.

We note, however,
that, since we are assuming knowledge of only the power spectra of each
component, the component separation algorithm has no a priori
information concerning cross correlations between any of the components.
In practice, a significant amount of cross correlation information should be available 
to the Planck mission from earlier observations, and so our assumed 
level of prior knowledge is, in this respect, mildly pessimistic.

\subsubsection{Reconstructed and residuals maps}

The component separation algorithm produces an estimate
$\hat{\mathbfss a}_{\ell m}$ of the harmonic coefficients of the
physical components at each mode. Let $a^{(p)}_{\ell m}$ be the true
harmonic coefficient of the $p$th physical component at the mode
$(\ell, m)$. This coefficient can be written
\begin{equation}
a^{(p)}_{\ell m} = \hat{a}^{(p)}_{\ell m} + \delta a^{(p)}_{\ell m},
\label{asum}
\end{equation}
where $\hat{a}^{(p)}_{\ell m}$ is the estimate of the mode obtained from the
component separation algorithm and $\delta a^{(p)}_{\ell m}$ is the
residual on this mode. The reconstructed map of the $p$th physical
component is given simply by performing an inverse spherical harmonic
transform on the reconstructed coefficients $\hat{a}^{(p)}_{\ell m}$
using the {\sc HEALPix} package. It is also of interest to construct a map
of the reconstruction residuals for each physical component. This can
be calculated in two equivalent ways. One can either simply subtract
the reconstructed map from the true input map, or one can perform
an inverse spherical harmonic transform of the residuals $\delta
a^{(p)}_{\ell m}$. We have verified that these two methods lead to
residuals maps that are consistent to within the machine precision.

\begin{figure*}
\begin{center}
\centerline{
\epsfig{
file=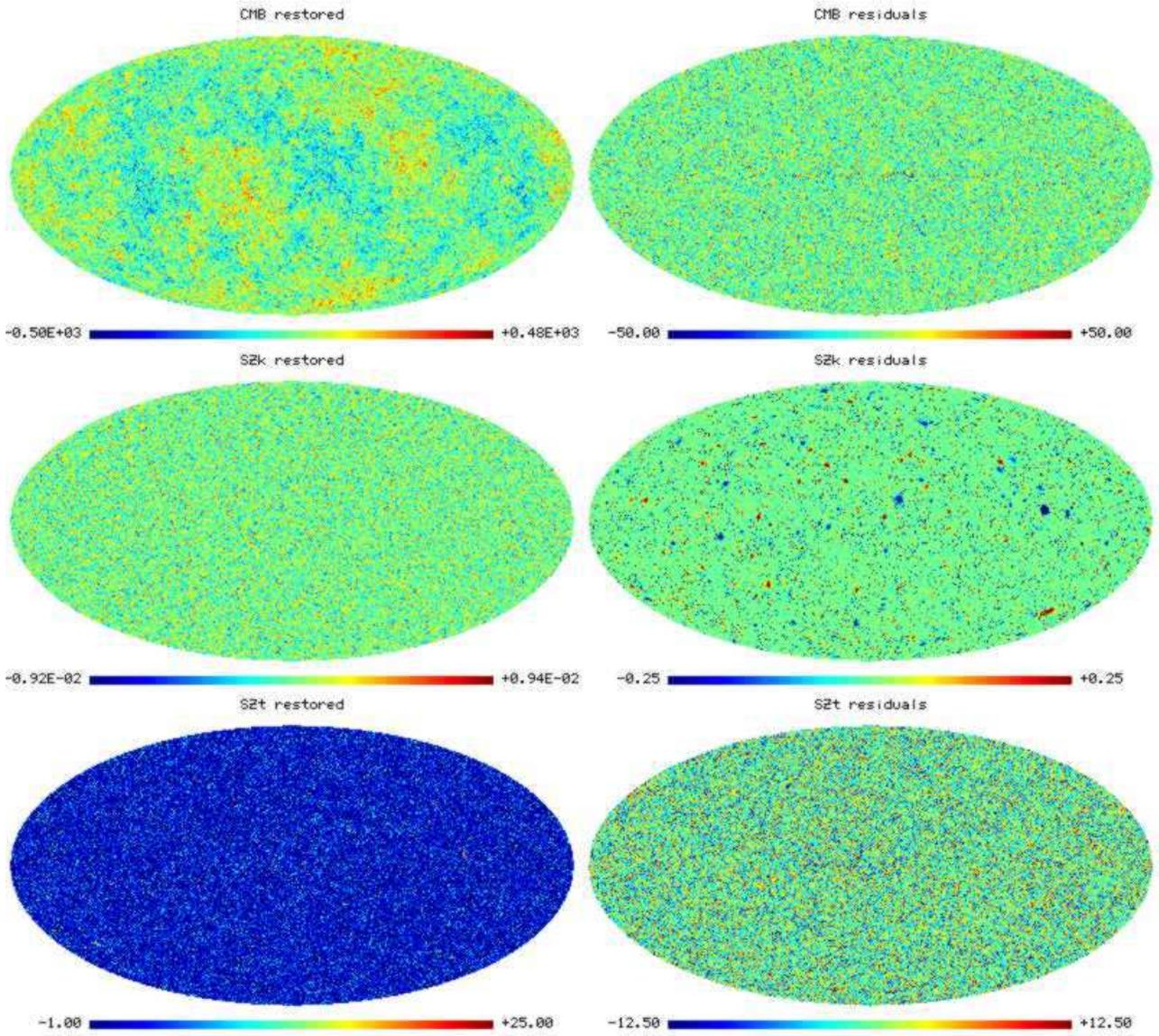,
width=17cm}}
%\centerline{
%\epsfig{
%file=CMB_restored_noicft.ps,
%angle=-90,width=8.5cm}
%\quad 
%\epsfig{
%file=CMB_res_noicft.ps,
%angle=-90,width=8.5cm}
%}
%\centerline{
%\epsfig{
%file=SZk_restored_noicft.ps,
%angle=-90,width=8.5cm}
%\quad 
%\epsfig{
%file=SZk_res_noicft.ps,
%angle=-90,width=8.5cm}
%}
%\centerline{
%\epsfig{
%file=SZt_restored_noicft.ps,
%angle=-90,width=8.5cm}
%\quad 
%\epsfig{
%file=SZt_res_noicft.ps,
%angle=-90,width=8.5cm}
%}
\caption{Reconstructions of the extragalactic components, namely
the primordial CMB, the kinetic SZ effect and the thermal
SZ effect (left-hand column). Also plotted are maps of
the corresponding reconstruction residuals for 
each component (right-hand column). All maps are plotted in units of $\mu$K.}
\label{fig4}
\end{center}
\end{figure*}

In Fig.\ref{fig4}, 
we plot the reconstructions and residuals of the extragalactic
components, namely the CMB and the kinetic and thermal SZ effects.
\begin{figure*}
\begin{center}
\centerline{
\epsfig{
file=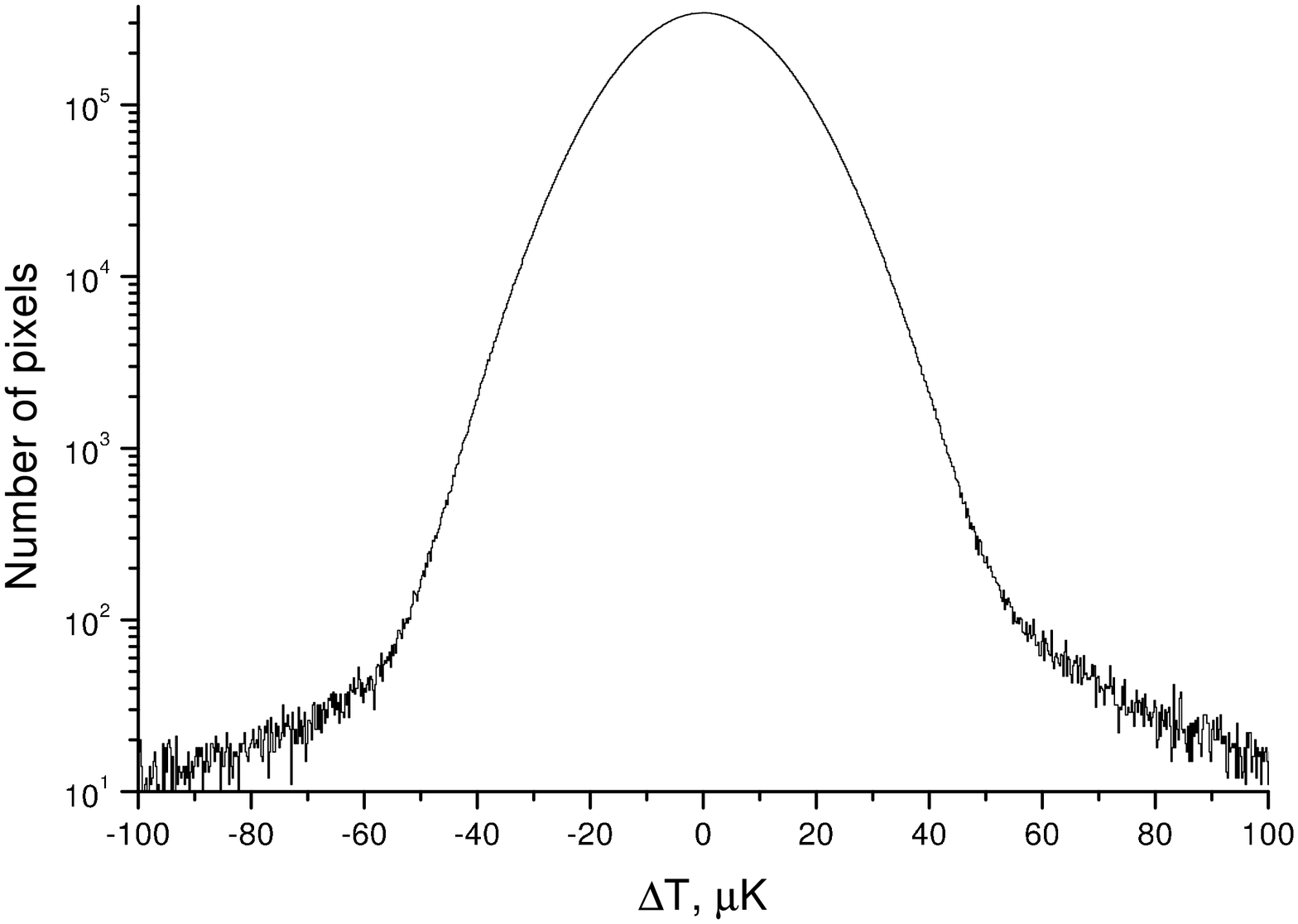,width=5.8cm}
\quad 
\epsfig{
file=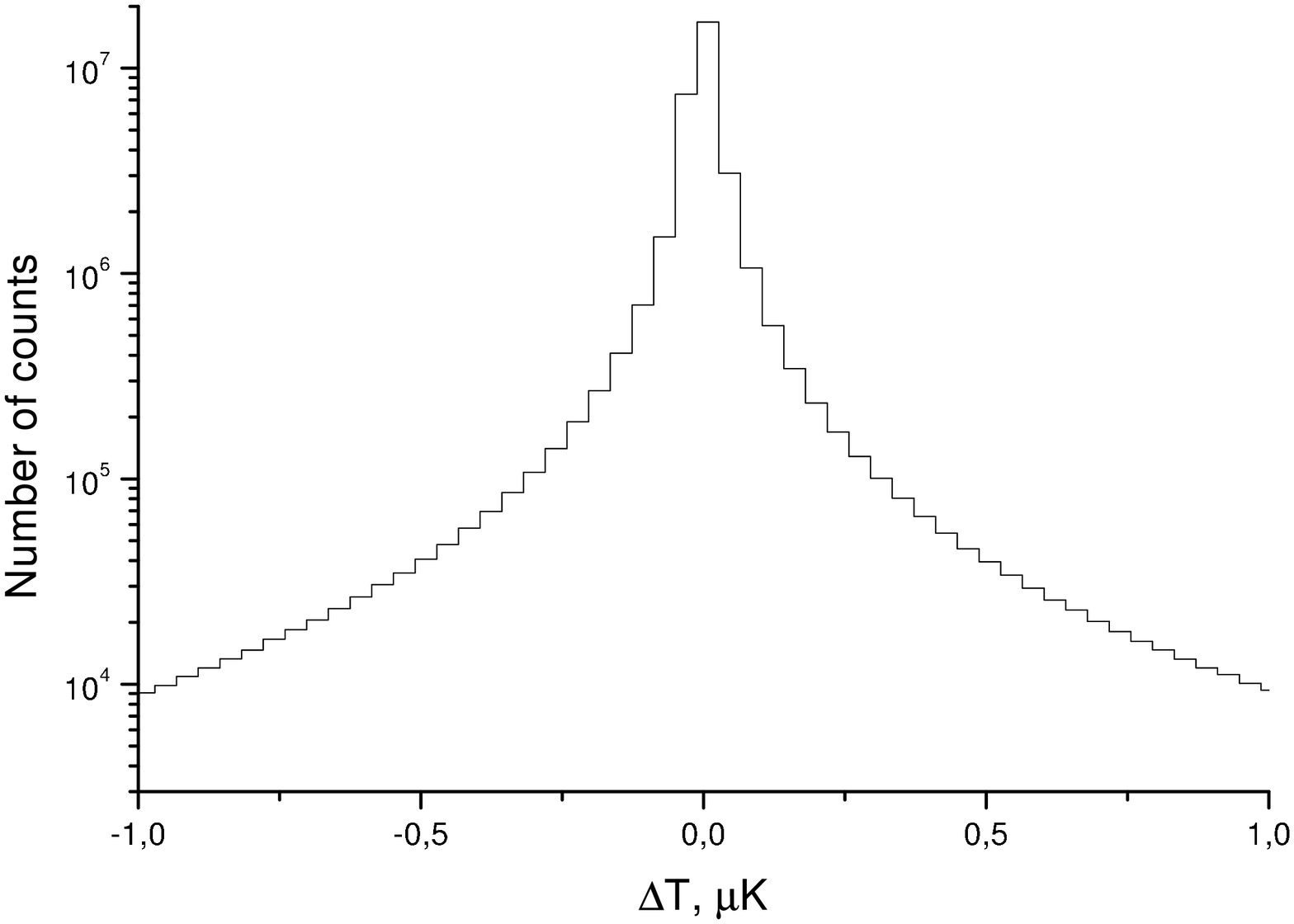,width=5.8cm}
\quad 
\epsfig{
file=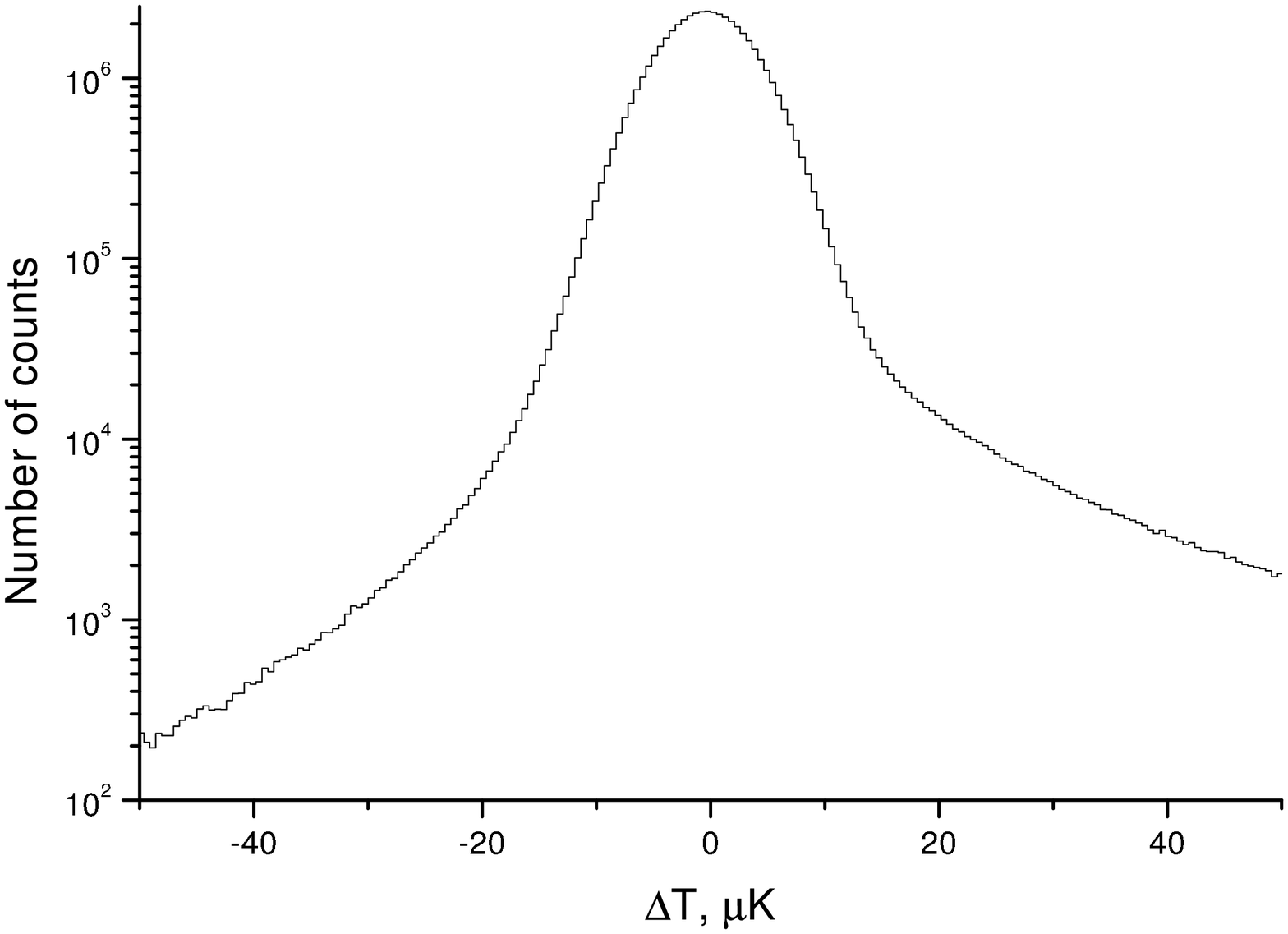,width=5.8cm}
}
\caption{Histograms of the residuals maps for primordial CMB (left), the 
kinetic SZ effect (centre) and the thermal SZ effect (right).}
\label{fig5}
\end{center}
\end{figure*}
\begin{figure*}
\begin{center}
\centerline{
\epsfig{
file=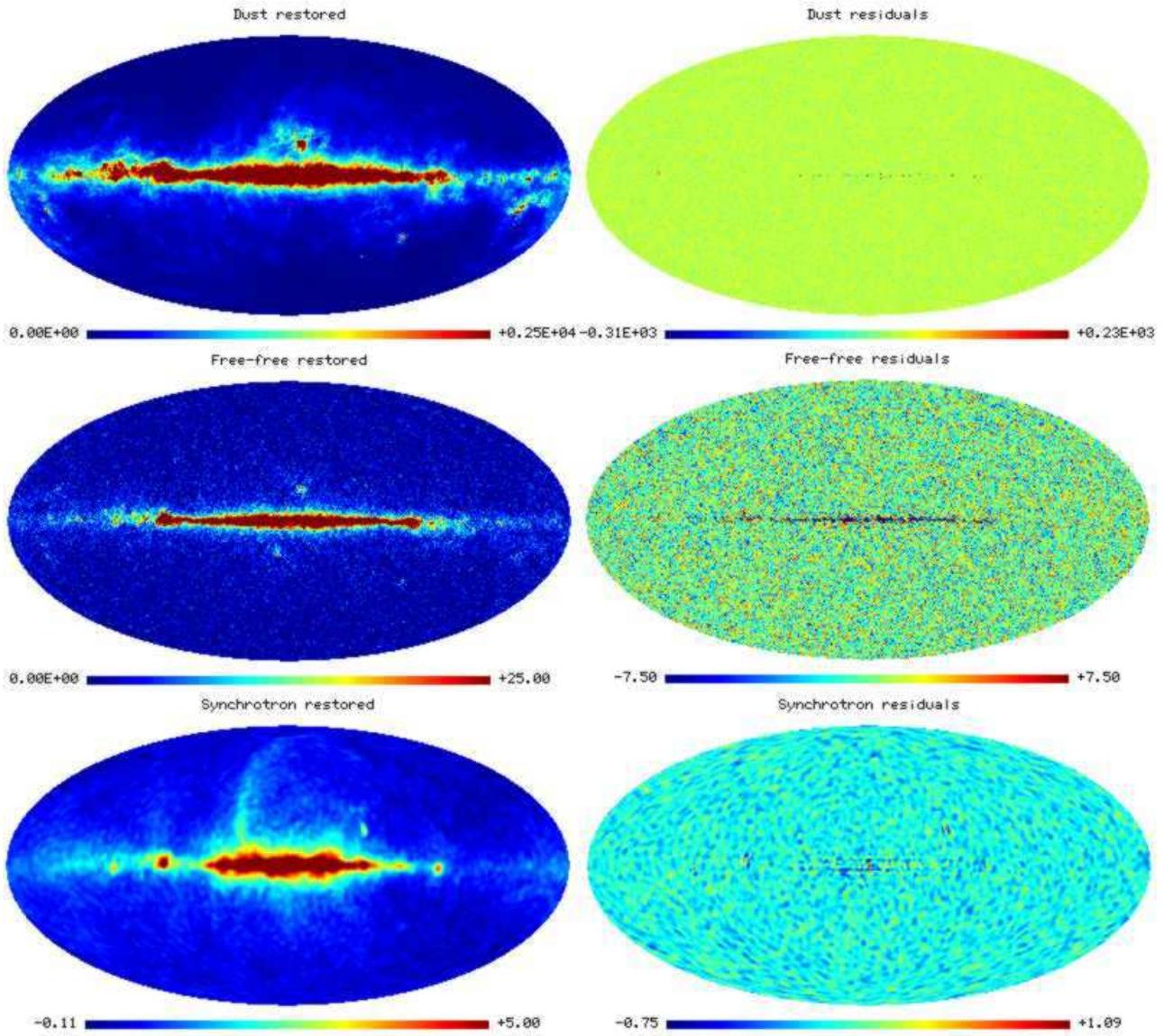,
width=17cm}}
%\centerline{
%\epsfig{
%file=Dust_restored_noicft.ps,
%angle=-90,width=8.5cm}
%\quad 
%\epsfig{
%file=Dust_res_noicft.ps,
%angle=-90,width=8.5cm}
%}
%\centerline{
%\epsfig{
%file=FF_restored_noicft.ps,
%angle=-90,width=8.5cm}
%\quad 
%\epsfig{
%file=FF_res_noicft.ps,
%angle=-90,width=8.5cm}
%}
%\centerline{
%\epsfig{
%file=Synch_restored_noicft.ps,
%angle=-90,width=8.5cm}
%\quad 
%\epsfig{
%file=Synch_res_noicft.ps,
%angle=-90,width=8.5cm}
%}
\caption{Reconstructions of the Galactic components, namely
the dust, free-free and synchrotron emission (left-hand column). 
Also plotted are maps of
the corresponding reconstruction residuals for 
each component (right-hand column).}
\label{fig6}
\end{center}
\end{figure*}
\begin{figure*}
\begin{center}
\centerline{
\epsfig{
file=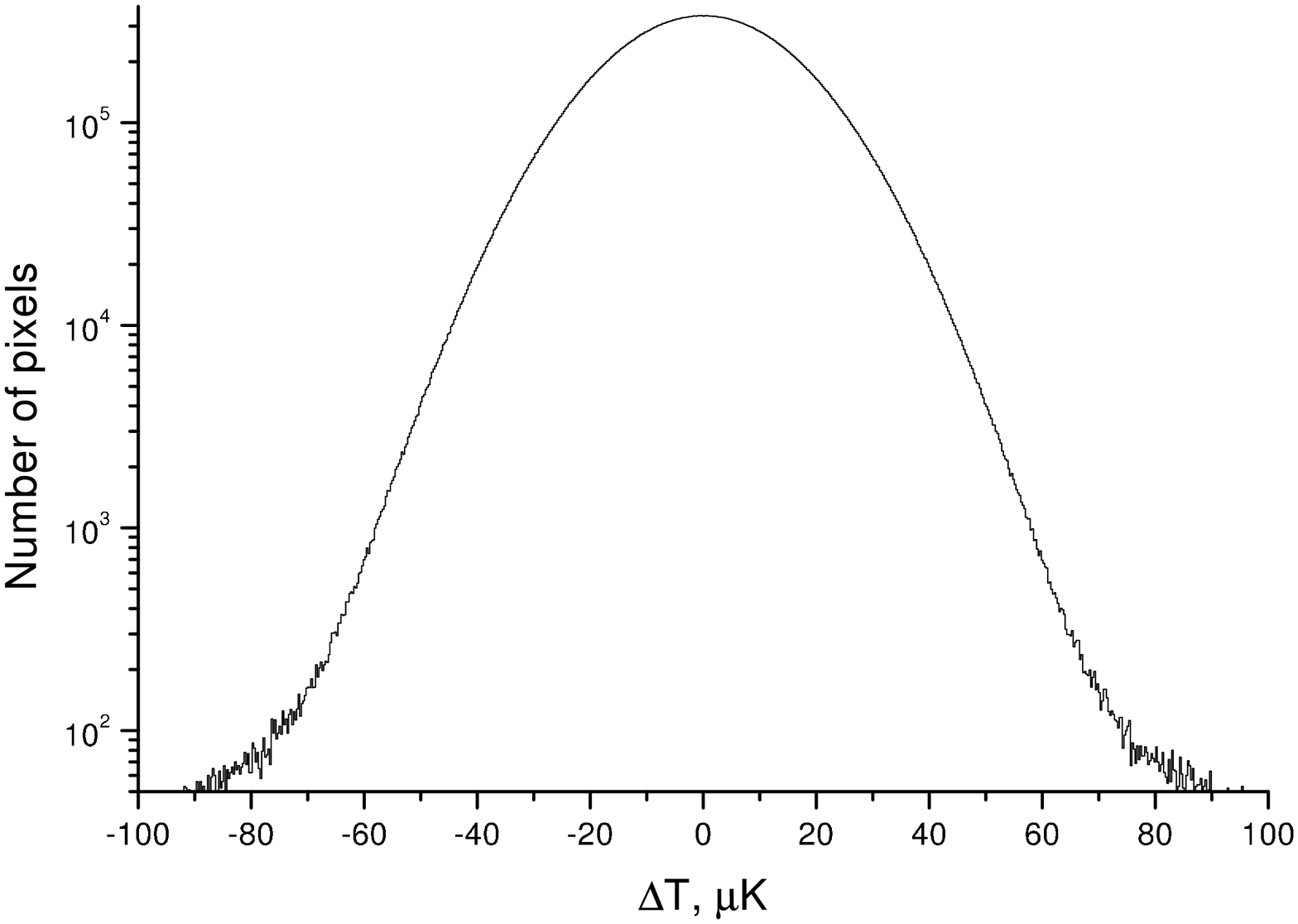,width=5.8cm}
\quad 
\epsfig{
file=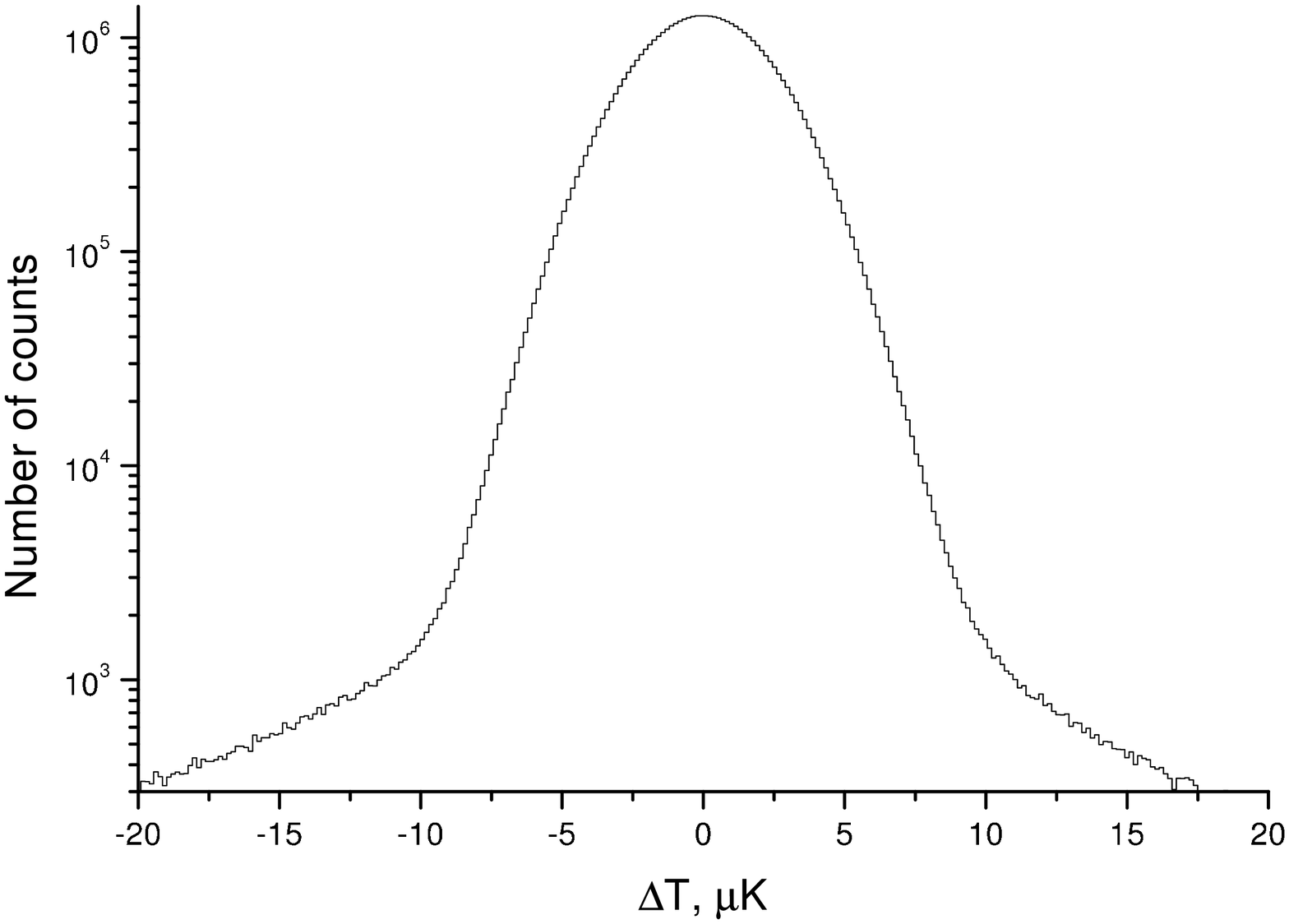,width=5.8cm}
\quad 
\epsfig{
file=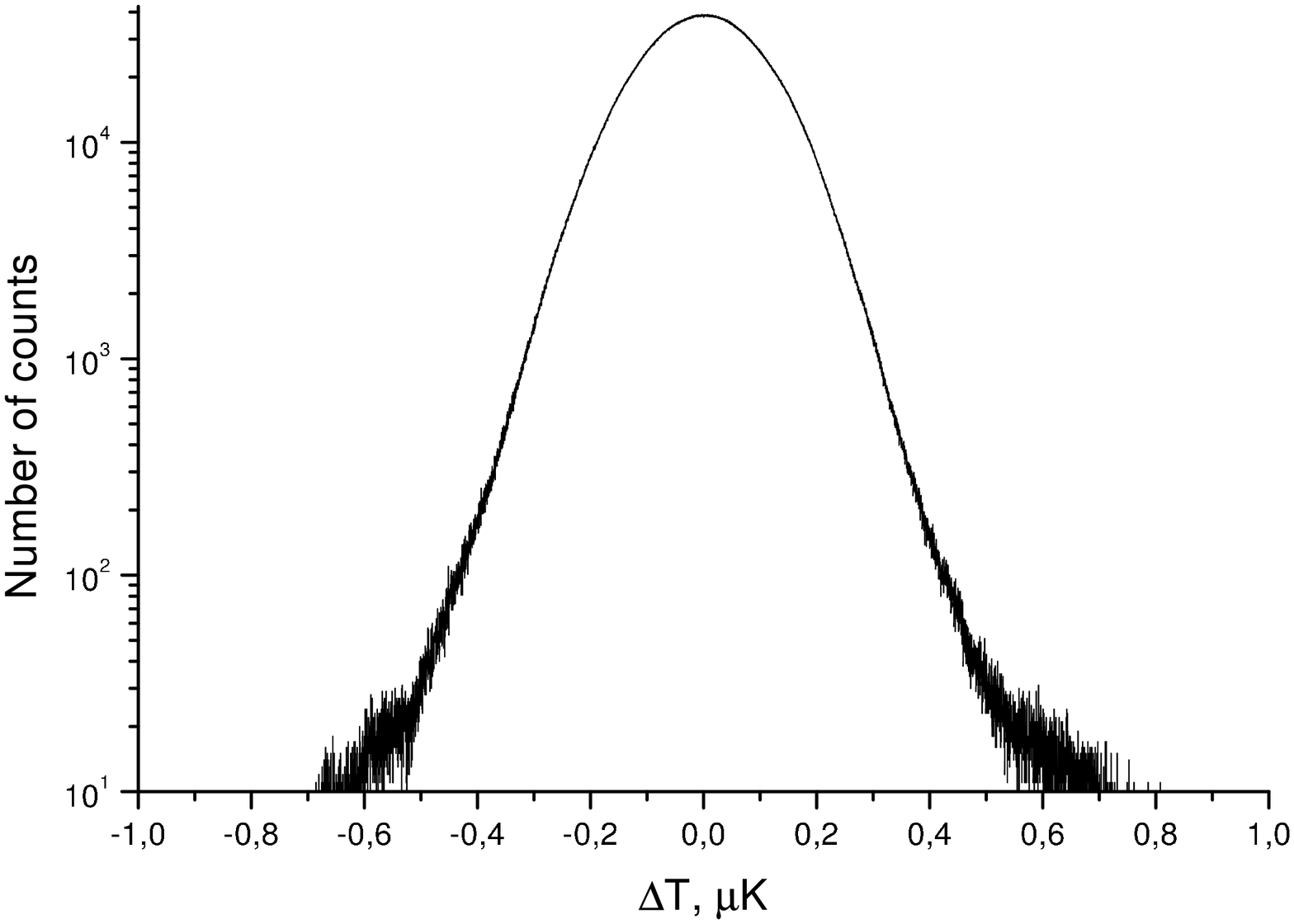,width=5.8cm}
}
\caption{Histograms of the residuals maps for emission from 
dust (left), free-free (centre) and synchrotron (right).}
\label{fig7}
\end{center}
\end{figure*}
We see that the CMB reconstruction is a faithful respresentation of
the input map shown in Fig.~\ref{fig1}. This is confirmed by
the CMB residuals map, which contains very little structure. 
Indeed, only a thin band of residual emission remains in the
CMB reconstruction in the very centre of the Galactic plane. 
Thus, the CMB map has been faithfully recovered over the whole sky, {\em
without} the need to perform a Galactic cut prior to the component
separation. It is also worth noting that the true input CMB map was
{\em not} smoothed before calculating the reconstruction residuals.
Fig.\ref{fig5} shows the distribution of pixel values in
the residuals map. This distribution is well approximated by a Gaussian
centred on zero with an rms of 12.3 $\mu$K, except for low-level 
wings corresponding to the very few pixels in the Galactic 
plane that contain residual
Galactic contamination. 

In the absence of a priori cross-correlation information, we see that
the recovery of the kinetic SZ effect is very poor, confirming the
findings of Paper I. This result is not surprising, since the kinetic
SZ effect has the same frequency spectrum
as the primordial CMB but, as shown in Figs~\ref{fig2} and \ref{fig3}, 
its rms signal is
3 orders of magnitude smaller than the CMB component. However, 
the thermal SZ effect has been successfully recovered, particularly in
the brighter clusters. The numbers of clusters recovered and accuracy
achieved, together with the resulting cosmological information that
could be obtained,
will be discussed in detail in a forthcoming paper.

In Fig.\ref{fig6}, we plot the reconstructions and residuals of the Galactic
components, namely dust, free-free and synchrotron emission. 
Once again we see that each component has been successfully recovered 
over an extremely wide dynamic range in pixel values.
In particular, the presence of strong emission from the
Galactic plane clearly invalidates the assumption of
uncorrelated harmonic modes, as discussed in section
\ref{compsep}. Nevertheless, this approximation does not appear to have
had a dramatic effect on the quality of the reconstructions.
The accuracy of the reconstructions is confirmed by
the corresponding residuals maps, which are mostly featureless
apart from some slight residual emission in the central Galactic
plane. The histograms of the pixel values in these residual maps
are shown in Fig.~\ref{fig7}. As for the CMB, each histogram is well
approximated by a Gaussian centred on zero, expect for low-level
wings corresponding to the few pixels in the Galactic plane that
contain residual contamination from other components. The parameters
of the best-fit Gaussian to each of the residuals histograms in 
Figs~\ref{fig5} \& \ref{fig7} are given in Table~\ref{table2}.
\begin{table}
\begin{center}
\caption{The mean $\mu$ and standard deviation $\sigma$ (in $\mu$K) of 
the best-fit Gaussian to
each of the residuals histograms shown in Figs~\ref{fig5} \& \ref{fig7}.}
\label{table2}
\begin{tabular}{lrr} \hline
Component &        $\mu$\phantom{a,}& $\sigma$\phantom{a,} \\ \hline
CMB              & 0.11 & 12.30 \\
Kinetic SZ       & 0.02 & 0.12 \\
Thermal SZ       & 0.62 & 4.30 \\
Dust             & 0.14 & 16.70 \\
Free-Free        & 0.08 & 2.40 \\
Synchrotron      & $3\times 10^{-4}$ & 0.11 \\ \hline
\end{tabular}
\end{center}
\end{table}
We note that the low values of $\sigma$ for the kinetic SZ effect
and Galactic synchrotron emission do not indicate that they are the
most accurately reconstructed, but simply that the level of emission
in these two components is very low at 300 GHz.

\subsubsection{Power spectra of reconstructed and residuals maps}

\begin{figure*}
\begin{center}
\centerline{
\epsfig{
file=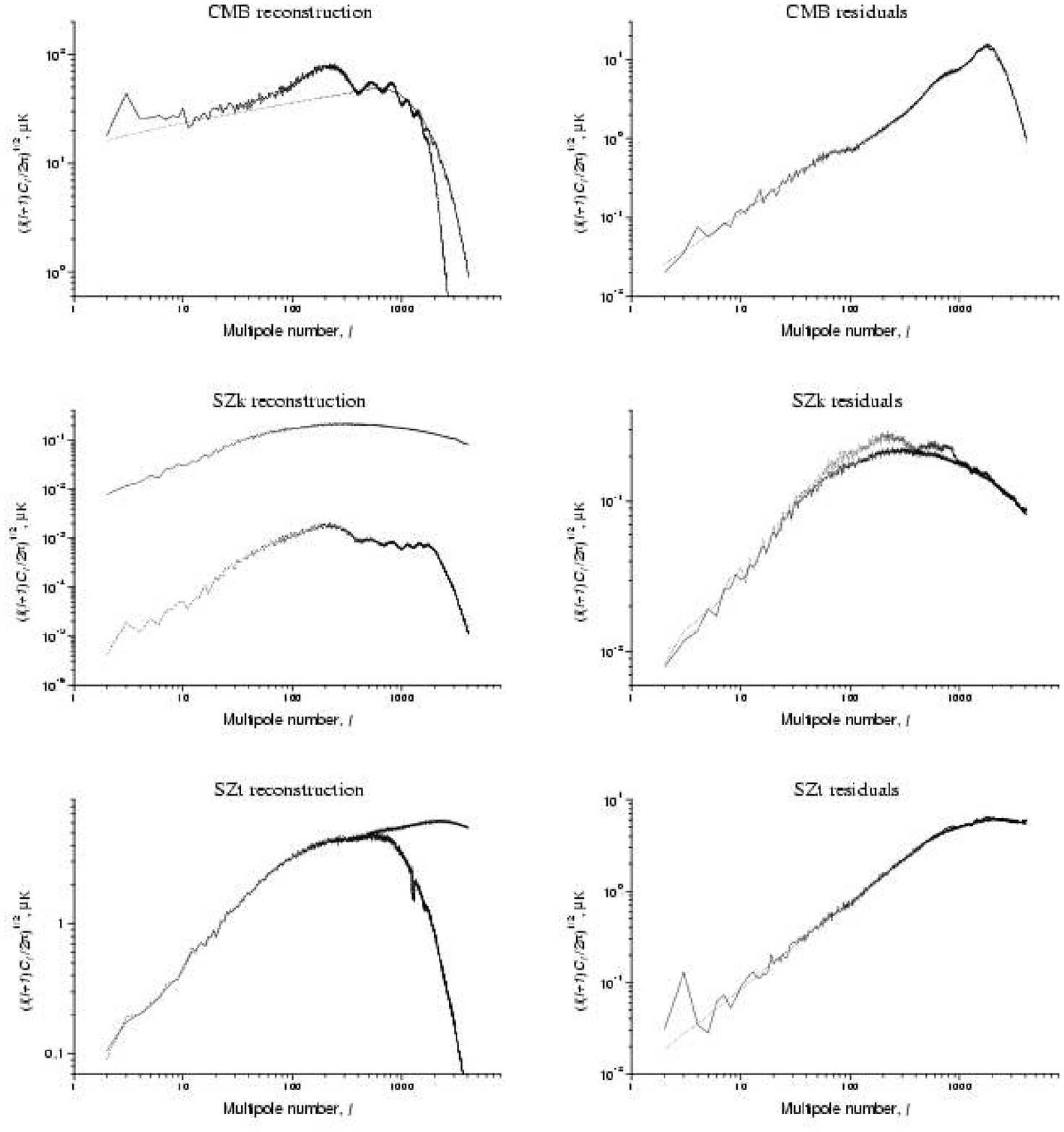,
width=16.0cm}}
%\centerline{
%\epsfig{
%file=fig8.eps,
%width=16.0cm}
%}
\caption{Power spectra of the reconstructed maps (dotted line) 
and input maps (solid line)
of the extragalactic components, namely CMB, kinetic SZ and thermal SZ
(left-hand column). The smooth dotted line in the CMB panel is the
assumed input CMB power spectrum supplied to the algorithm.
Also plotted are the power spectra of the residuals
maps (solid line) and the predicted residuals power spectra (dotted line)
for each component (right-hand column).}
\label{fig8}
\end{center}
\end{figure*}

\begin{figure*}
\begin{center}
\centerline{
\epsfig{
file=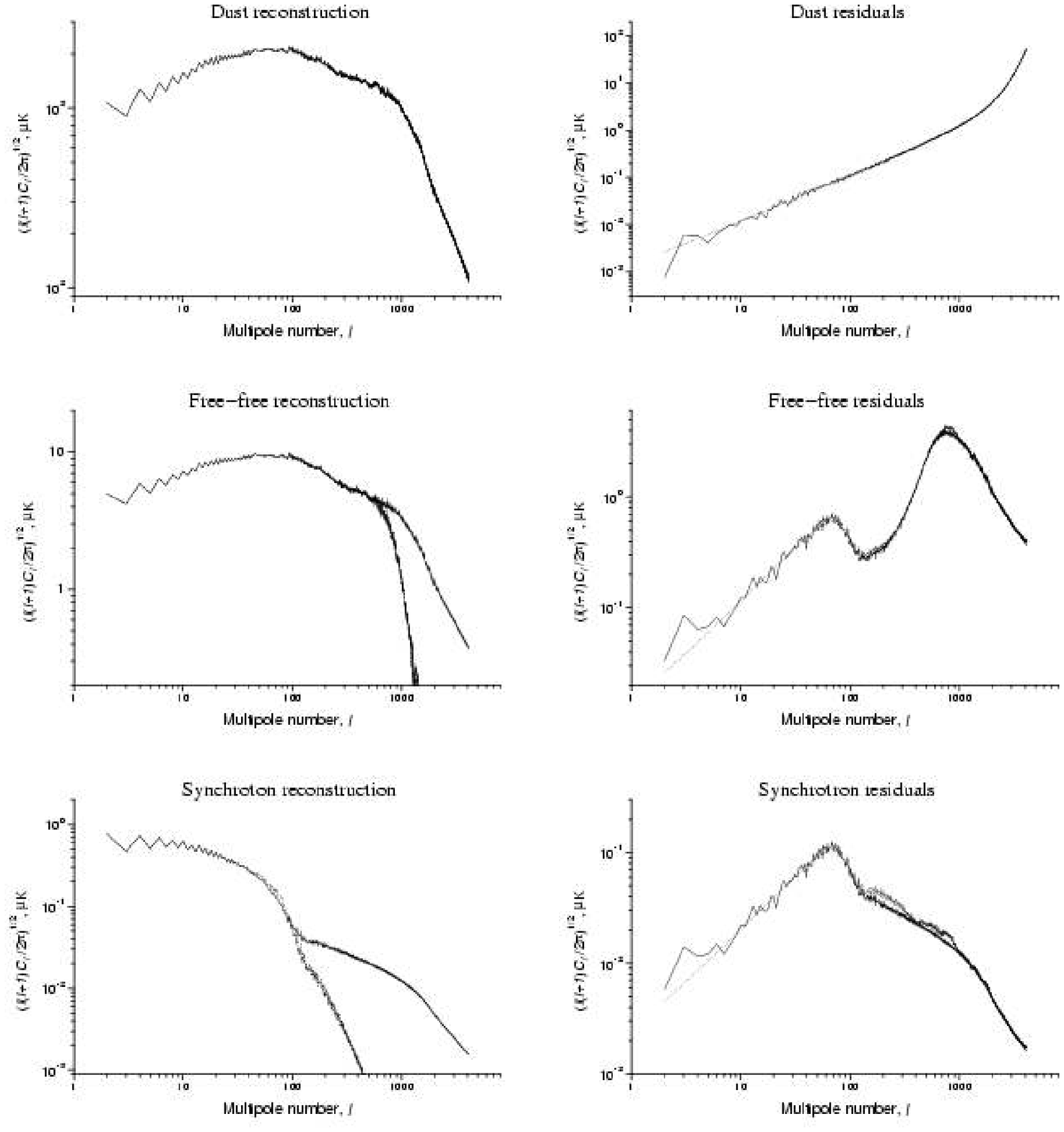,
width=16.0cm}}
%\centerline{
%\epsfig{
%file=fig9.eps,
%width=16.0cm}
%}
\caption{As in Fig.~\ref{fig8}, but for the Galactic components dust, free-free
and synchrotron.}
\label{fig9}
\end{center}
\end{figure*}

In the left-hand columns of Figs~\ref{fig8} \& \ref{fig9} , we plot the 
power spectrum of the
reconstructed map for each physical
component (dotted line), together with the power spectrum of the true
input map (solid line).
For the $p$th physical component, the power spectrum of
the reconstructed map is given simply by
\begin{equation}
\hat{C}^{(p)}_{\ell}  \equiv  \frac{1}{2\ell +1} \sum_{m=-\ell}^\ell 
|\hat{a}^{(p)}_{\ell m}|^2,
\label{clsimp}
\end{equation}
In the right-hand column of Figs~\ref{fig8} \& \ref{fig9}, we
plot the power spectrum of the residuals
map for each component (solid line), which is given by
\begin{equation}
\delta C^{(p)}_{\ell}  
\equiv  \frac{1}{2\ell +1} \sum_{m=-\ell}^\ell |\delta a^{(p)}_{\ell m}|^2.
\label{dcsimp}
\end{equation}
Of course, in practice the true residuals  $\delta a_{\ell m}$ would
not be available. Nevertheless, we can estimate the
residuals power spectrum by calculating the Hessian
matrix ${\mathbfss H}_{\ell m}$ of the posterior distribution at its peak.
As discussed in section \ref{compsep}, this provides an estimate of the
standard error on the reconstructed harmonic modes. From
(\ref{hesserr}), we see that for the $p$th physical component
$|\Delta a^{(p)}_{\ell m}|^2 \equiv \langle |\delta a^{(p)}_{\ell m}|^2 \rangle$ is
given by the $p$th diagonal entry of the matrix 
${\mathbfss L}_\ell {\mathbfss H}^{-1}_{\ell m} {\mathbfss
L}^\dagger_\ell$. Thus, our estimator of the residuals power
spectrum is simply
\begin{equation}
\widehat{\delta C}^{(p)}_\ell = \frac{1}{2\ell +1} \sum_{m=-\ell}^\ell |\Delta
a^{(p)}_{\ell m}|^2, 
\label{dchat}
\end{equation}
which is unbiassed in the sense that
$\widehat{\delta C}^{(p)}_\ell = \langle \delta C^{(p)}_\ell
\rangle$, where the averages are over realisations of the
instrumental noise and maps of the physical components. The
estimated residuals power spectrum for each component is plotted
as the dotted line in the right-hand columns of Figs~\ref{fig8} \& \ref{fig9}.

Fig.~\ref{fig8} shows the power spectra for the extragalactic
components, namely the CMB and kinetic and thermal SZ effects.
For the CMB, in addition to the power spectra of the
input and reconstructed maps, we also plot the simple a priori form of
the power spectrum assumed by the component separation algorithm.
This assumed power spectrum contains no acoustic peak structure, so as
not to prejudice the reconstruction. We see, however, that the power
spectrum of the reconstructed map matches that of the input map up to
$\ell \approx 2000$. In particular, the first 5 acoustic peaks have
been accurately recovered, and there is some evidence of the presence
of a 6th peak in the recovered power spectrum, although the
amplitude is significantly underestimated. Indeed, beyond $\ell
\approx 2000$, the power spectrum of the reconstructed map consistently
underestimates that of the true map. This is effect is due to the fact
that (\ref{clsimp}) is a biassed estimator of the true power spectrum. 
Nevetheless, as
discussed below, we can easily construct an
unbiassed estimator. From Fig.~\ref{fig8}, however, we see that our
estimator for the power spectrum of the CMB residuals is already unbiassed,
as we expected. Thus the MEM algorithm provides a good estimate of the
statistics of the reconstruction errors.

The power spectra of the reconstructed kinetic SZ map severely
underestimates the true power spectrum, as we might have
expected. Indeed, one can see that the `recovered' kinetic SZ map has
a power spectrum which clearly mirrors that of the CMB, which suggests
that any structure present in the kinetic SZ `reconstruction' is dominated by
contamination by primordial CMB. Neverthelss, once again the 
estimated residuals
power spectrum is a good approximation to the power spectrum of the
true residuals map. In practice this would provide a robust indication
of the poor quality of the kinetic SZ reconstruction.

For the thermal SZ effect, the power spectrum of the reconstructed map
matches that of the input map up to $\ell \approx 1000$, beyond which
the true spectrum is underestimated. This is again is result of 
(\ref{clsimp}) being an biassed estimator. We see, however, that
the predicted and actual residuals power spectra are in close
agreement for all $\ell$ values.

The power spectra of the reconstructed maps and residuals of the
Galactic components are shown in Fig.~\ref{fig9}. Qualitatively, we see
similar behaviour to the observed for the extragalactic components,
with the exception that the power spectrum of the input dust map is 
accurately recovered for all $\ell$. For the free-free and synchrotron
components, the underestimation of the true power spectrum is again
due to the fact that (\ref{clsimp} is a biassed estimator.
In each case, we again note that
the estimated and actual residuals power spectrum agree well. The
accuracy of this error estimation enables one to define an unbiassed 
estimator of the true input power spectrum.

\subsubsection{An unbiased power spectrum estimator}

We see from the left-hand columns of Figs~\ref{fig8} \& \ref{fig9}, 
that the power spectra of the reconstructed
maps generally underestimate the true input power spectra at large $\ell$.
Indeed, it is easy to show that the estimator (\ref{clsimp}) is biassed.
From (\ref{asum}), we have $\hat{a}^{(p)}_{\ell m}=
a^{(p)}_{\ell m}-\delta a^{(p)}_{\ell m}$, and on inserting this
expression into (\ref{clsimp}), we see that the expectation
value of this estimator is given by
\[
\langle \hat{C}^{(p)}_{\ell} \rangle  
 =  \frac{1}{2\ell +1} \sum_{m=-\ell}^\ell 
\langle |a^{(p)}_{\ell m}|^2 \rangle + 
\langle |\delta a^{(p)}_{\ell m}|^2\rangle
-2\langle a^{*(p)}_{\ell m}\delta a^{(p)}_{\ell m}\rangle 
\]
where the ensemble average is again taken over realisations of the
instrumental noise and of the physical components.
Using (\ref{asum}) again, we can write
\begin{equation}
\langle a^{*(p)}_{\ell m}\delta a^{(p)}_{\ell m}\rangle
%& = & \langle (\hat{a}^{*(p)}_{\ell m}+ \delta a^{*(p)}_{\ell m})
%\delta a^{(p)}_{\ell m}\rangle \nonumber \\
 =  \langle \hat{a}^{*(p)}_{\ell m}\delta a^{(p)}_{\ell m}\rangle 
+ \langle |\delta a^{(p)}_{\ell m}|^2 \rangle. \label{inter} 
\end{equation}
If we assume that the reconstruction errors on each mode 
are not correlated with the
reconstructed values (which is reasonable given the
featureless nature of the residuals maps), the first term on the
right-hand side of (\ref{inter}) disappears, and so we obtain
\begin{eqnarray}
\langle \hat{C}^{(p)}_{\ell} \rangle  
& = & \frac{1}{2\ell +1} \sum_{m=-\ell}^\ell 
\langle |a^{(p)}_{\ell m}|^2 \rangle
- 
%\frac{1}{2\ell +1} \sum_{m=-\ell}^\ell  
\langle |\delta 
a^{(p)}_{\ell m}|^2 \rangle, \nonumber \\
& = & \langle C^{(p)}_\ell \rangle - \langle \delta C^{(p)}_\ell
\rangle, \nonumber \\
& = & \langle C^{(p)}_\ell \rangle 
\left(1-\frac{\langle \delta C^{(p)}_\ell \rangle}
{\langle C^{(p)}_\ell \rangle}\right).
\label{cbias}
\end{eqnarray}
Thus, (\ref{clsimp}) is a biassed estimator, 
which underestimates the
true power spectrum of the $p$th component
on angular scales where the residuals are
large. 

By analogy with the approach developed by Bouchet \& Gispert (1999)
and Knox (1995), we denote the quantity in brackets in (\ref{cbias}) as the
{\em quality factor} $Q^{(p)}_\ell$. We see immediately that
an unbiassed estimator of the power spectrum is given by simply by 
\begin{equation}
\hat{\cal C}^{(p)}_\ell  = \hat{C}^{(p)}_\ell
\left(1-\frac{\widehat{\delta C}^{(p)}_\ell}{\langle C^{(p)}_\ell
\rangle}\right)^{-1},
\label{unbias}
\end{equation}
where $\widehat{\delta C}^{(p)}_\ell$ is given by (\ref{dchat}). 
The variance of this estimator is given by
\begin{equation}
V \left[\hat{\cal C}^{(p)}_\ell \right]  = \frac{2}{2\ell +1} 
\left(\frac{\langle C^{(p)}_\ell \rangle}{Q^{(p)}_\ell}\right)^2.
\label{varunbias}
\end{equation}
which illustrates the uncertainty added by the foreground removal to 
cosmic variance.

\begin{figure}
\begin{center}
\centerline{
\epsfig{file=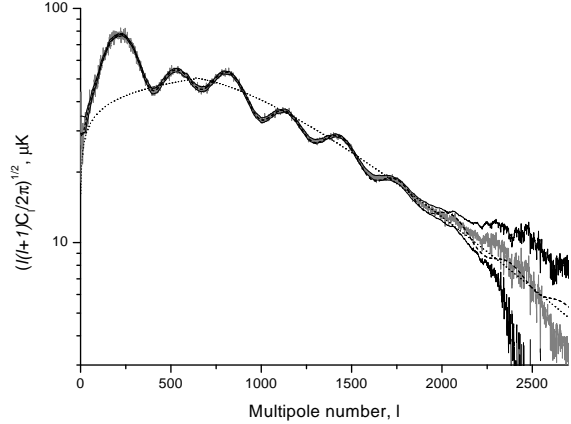,
width=9.5cm}}
\caption{The power spectrum of the input CMB map (dashed line) compared
with the unbiassed estimator (\ref{unbias}) of the power spectrum obtained
from the CMB reconstruction (grey solid line). Also plotted are the 
1-$\sigma$ error limits. The smooth dotted line shows the
power spectrum of the CMB assumed as our prior 
in the component separation process.}
\label{fig10}
\end{center}
\end{figure}

As an illustration, in Fig.~\ref{fig10} we plot the unbiassed
estimator of the CMB power spectrum against the true power spectrum of
the input map. 
We see from the plot that the estimated power spectrum 
is now fully consistent with the input power
spectrum out to large $\ell$ values.
In particular, the heights and positions of the 5th and 6th acoustic
peaks are now accurately recovered. Beyond the 6th peak, however, there is 
no evidence for further oscillatory structure in the power
spectrum. At high $\ell$, the overall level of power is recovered correctly, 
but this is simply a result of our
assumed power spectrum (the smooth dotted line in Fig.~\ref{fig10})
having the appropriate normalisation in this region.

\subsection{Full prior covariance information}

As an indication of the best results one could reasonably expect, we
now consider the case in which, in addition to the 
azimuthally-averaged power spectra of each input component, we also
assume knowledge of the azimuthally-averaged cross power spectra between
components. These provide some limited cross-correlation information between
the components at each spherical harmonic mode. Moreover, in this
case, we also provide the algorithm with the true power spectrum of the CMB
input map, with its full acoustic peaks structure, rather than the
simple functional form assumed above.

Most of the resulting reconstructed maps of components
are found to be very similar to those
obtained in the previous section, obtained using no cross-correlation
information. We therefore do not reproduce them here in full, but list
only the rms of the residuals maps in Table~\ref{table3}, which we see are 
very similar to those listed in Table~\ref{table2}.
\begin{table}
\begin{center}
\caption{The mean $\mu$ and standard deviation $\sigma$ (in $\mu$K) of 
the best-fit Gaussian to the residuals histograms for full a priori
correlation information.}
\label{table3}
\begin{tabular}{lrr} \hline
Component &        $\mu$\phantom{a,}& $\sigma$\phantom{a,} \\ \hline
CMB              & 0.09 & 12.18 \\
Kinetic SZ       & 0.02 & 0.09 \\
Thermal SZ       & 0.60 & 4.50 \\
Dust             & 0.14 & 16.70 \\
Free-Free        & 0.05 & 1.58 \\Synchrotron      & $5\times 10^{-4}$ & 0.10 \\ \hline
\end{tabular}
\end{center}
\end{table}

\begin{figure}
\begin{center}
\centerline{
\epsfig{file=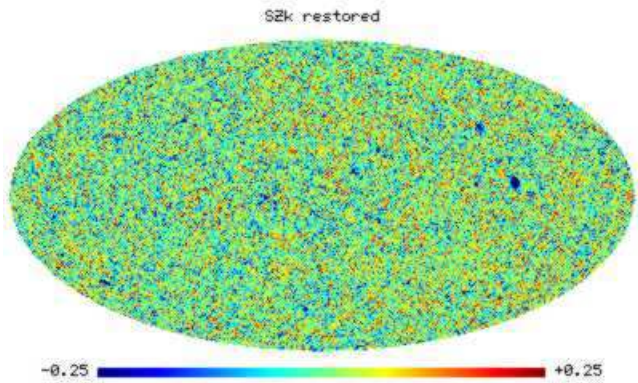,
width=8.5cm}}
%\centerline{
%\epsfig{file=SZk_restored_icft.ps,
%angle=-90,width=8.5cm}}
\caption{Reconstructed map of the kinetic SZ effect.}
\label{fig11}
\end{center}
\end{figure}

In this section, we instead concentrate on the few qualitative
differences in the reconstructions that occur due to the inclusion 
of cross-correlation information.
The most striking difference occurs in the reconstruction of the
kinetic SZ component. As shown in the previous section, in the absence
of cross-correlation information, the kinetic SZ effect is essential
not recovered at all. Indeed, from Fig.~\ref{fig8}, we see that the low
level emission in the kinetic SZ `reconstruction' is residual
contamination from the primordial CMB component. If
some cross-correlation information is included, however, we see in
Fig.~\ref{fig11}, that the reconstruction of the kinetic SZ somewhat
improved.
\begin{figure}
\begin{center}
\centerline{
\epsfig{file=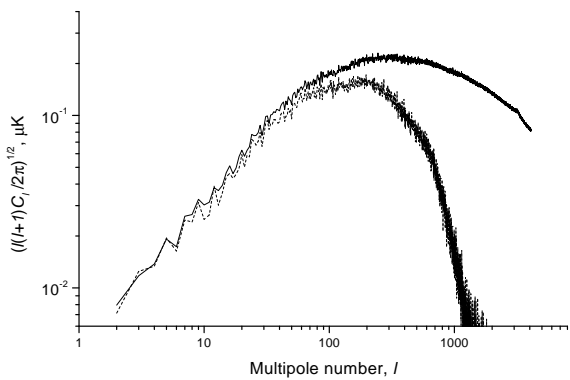,width=8.5cm}}
\caption{Power spectrum of reconstructed  map of kinetic SZ effect (dotted 
line) compared to the power spectrum of the input map (solid line).}
\label{fig12}
\end{center}
\end{figure}
In particular, as found in Paper I, the kinetic SZ effect is 
recovered to reasonable accuracy
only in those clusters possessing very large thermal
SZ effects. This clearly shows that the incorporation of
cross-correlation information between the two SZ effects enhances
the kinetic SZ reconstruction. The improved  quality of the
reconstruction is also illustrated by its power spectrum, which is
shown in Fig.~\ref{fig12} (together with the true power spectrum of the
input map). We see that at least up to
$\ell \approx 100$, the power in the reconstruction matches that of
the input map. This is a significant improvement on the power spectrum
of the reconstruction obtained using no cross-correlation information
shown in Fig.~\ref{fig8}.

\begin{figure}
\begin{center}
\centerline{
\epsfig{file=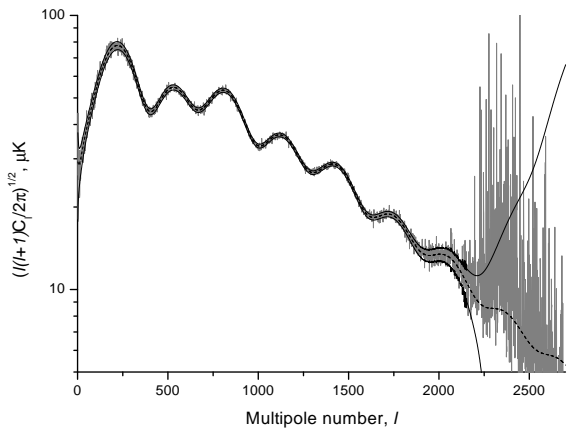,width=9.5cm}}
\caption{The power spectrum of the input CMB map (dashed line) compared
with the unbiassed estimator (\ref{unbias}) of the power spectrum obtained
from the CMB reconstruction (grey solid line). Also plotted are the 
1-$\sigma$ error limits.}
\label{fig13}
\end{center}
\end{figure}

The other major qualitative difference in the reconstruction concerns
the recovery of the acoustic peaks in the CMB power spectrum. In
Fig.~\ref{fig13} we plot the unbiassed estimator (as discussed above) 
obtained from the reconstruction. We also plot the true power spectrum 
of the input CMB
map and the 1-$\sigma$ error limits given by the square-root of the
variance given in (\ref{varunbias}).
To avoid inevitable scatter in the 1-$\sigma$ error limits at large $\ell$, 
$Q^{(p)}_\ell$ was approximated by the exponential function 
for $\ell \geq$2200.

Comparing Figs~\ref{fig10} and \ref{fig13}, we observe qualitative differences
at high-$\ell$. Most importantly, we note that more of the acoustic peaks
structure present in the true power spectrum is recovered.
In particular, the 7th acoustic peak is now reproduced
with its correct
height and position. Beyond the the 7th peak, however, error in our
simple unbiassed estimator becomes too large to discern any further
detailed structure in the power spectrum.
This result is in a good agreement with the 
analytical predictions by Bouchet \& Gispert (1999) based on Wiener
filtering of Gaussian foregrounds.

\section{Discussion}

In this paper, we have performed an all-sky component separation 
over the whole sky for the Planck mission at full angular resolution.
The technique employed is an harmonic-space maximum-entropy
method, which performs a mode-by-mode reconstruction of the spherical
harmonic coefficients of the component maps. Assuming knowledge only
of the power spectra of the non-CMB components, the algorithm produces
faithful reconstructions of the input maps, {\em without} the need to
apply a Galactic cut. Indeed, it is the MEM prior that allows our
component separation algorithm to cope with such a large dynamics
range in pixel values. The only component not reconstructed is
the kinetic SZ effect. We also assumed that the instrumental pixel noise
is homogeneous and uncorrelated. However, non-homogeneous uncorrelated pixel
noise can be easily included into the analysis. Preliminary
investigations of simulations for more realistic Planck observing
strategies suggest that the quality of the reconstructions are not
significantly reduced, provided one obtains reasonable sky coverge.
A detailed analysis of the effects of scanning strategy on the
accuracy of component separation will be presented in a forthcoming
paper.

The power spectra of the reconstructions match those of the input
maps (expect for the kinetic SZ)  on scales where the signal dominates 
the instrumental noise. At high $\ell$ values, however, the power
spectrum
of the reconstruction consistently underestimates the power spectrum of the
true map. Indeed, it is easily shown that this estimator is biassed.
Nevertheless, the component separation algorithm also produces an
accurate predicted power spectra for the
reconstruction residuals, which allows the construction of an
unbiassed power spectrum estimator. This accurately recovers the input
power spectrum at all measured $\ell$.

The effect of including a priori cross-correlation information is also
studied. The main effect is to improve somewhat the recovery of the 
kinetic SZ effect in clusters possessing a large thermal SZ effect.
The assumed level of prior knowledge of the CMB power spectrum also affects
the CMB reconstruction, but only at high $\ell$. In particular, we
find that, if ones assumes a CMB power spectrum with no acoustic
peaks, the final reconstructed spectrum still recovers the oscillatory
structure out to the 6th acoustic peak. If, however, the true CMB
power spectrum is assumed in the component separation process, the
recovered spectrum of the CMB contains the first 7 acoustic peaks
with the correct heights and positions.
It is also noted that some care must be taken in interpreting
the recovered CMB power spectrum  beyond 
$\ell \approx 2000$. In this region, the recovered spectrum is dominated by the
assumed spectrum. 

The simulations used in this paper did not contain a contribution from
extragalactic point sources. As shown in Hobson et al. (1998),
however, the MEM method can accommodate such a components. Moreover,
as discussed by Vielva et al. (2001), the MEM algorithm can be
combined with a mexican hat wavelet approach to detecting and removing
point sources, which leads to a significant reduction in the residual
point source contamination of the other physical components.
Recently, Cayon et al. (2001) have presented a consistent method
for defining the mexican hat wavelet on a sphere, which can be
combined with the algorithm presented here to separate out point
source emission in all-sky maps. This will be discussed in detail
in a forthcoming paper. In a future publication, we will also present
a detailed statistical analysis of the recovery of the kinetic and
thermal SZ effects in clusters, and discuss its cosmological implications.

\section{Acknowledgements}

We thank P.~Fosalba and G.~Giardino for providing the synthetic maps of the
synchrotron template and spectral index, and R\"udiger Kneissl for
providing the simulations of the thermal and kinetic SZ effects. 
We are also grateful to Daniel 
Mortlock for several illuminating discussions regarding orthonormal
bases on the cut-sky. We acknowledge the {\sc HEALPix} collaboration
for providing the pixelisation scheme used in this work.

\label{lastpage}

\end{document}